\documentclass[a4paper,11pt]{article}

\pdfoutput=1 

\usepackage{jheppub} 
\usepackage[T1]{fontenc} 

\usepackage{graphicx} 
\usepackage[utf8]{inputenc}
\usepackage{comment}
\usepackage{url}
\usepackage{hyperref}
\usepackage{tikz}
\usetikzlibrary{decorations.pathmorphing,braids,knots}
\usetikzlibrary{decorations.markings}
\usetikzlibrary{arrows.meta}
\usepackage{amsmath,amssymb,tikz-cd}
\usepackage{subcaption}
\usepackage{braket}
\usepackage{amsthm}
\usepackage[dvipsnames]{xcolor}
\tikzset{snake it/.style={decorate, decoration=snake}}
\tikzset{coil it/.style={decorate, decoration={coil,aspect=0.6,segment length=2mm,amplitude=1mm}}}

\definecolor{red}{rgb}{1,0,0}
\long\def\comment#1{}
\usepackage{chessfss}
\pdfoutput=1
\pdfpageattr {/Group << /S /Transparency /I true /CS /DeviceRGB>>}

\hypersetup{colorlinks=true,linkcolor=blue,citecolor=blue}
\usepackage{amsfonts}
\usepackage{bm}
\usepackage{float}

\usepackage{scalefnt}
\usepackage{wrapfig}
\usepackage{todonotes}
\usepackage{fancybox}
\usepackage{multirow}
\usepackage{adjustbox}
\usepackage{circuitikz}
\usepackage[toc,page]{appendix}

\usepackage{upgreek}

\usepackage[none]{hyphenat}
\title{On the monodromy of KZ-connections with irregular singularities}
\author[\dag]{Xia Gu,}
\author[\ast\dag]{Babak Haghighat,}
\author[\sharp]{Pavel Putrov}
\affiliation[\dag]{Yanqi Lake Beijing Institute of Mathematical Sciences and Applications (BIMSA), Huairou District, Beijing 101408, P. R. China}
\affiliation[\ast]{Yau Mathematical Sciences Center, Tsinghua University, Beijing, 100084, China}
\affiliation[\sharp]{ The Abdus Salam International Centre for Theoretical Physics, \\
 Strada Costiera 11, 34151 Trieste, Italy}
\abstract{We study Knizhnik-Zamolodchikov (KZ) connection in the presence of irregular singularities, that is, poles of higher order. We consider both the case of a universal connection and the case when it is associated with a specific simple Lie algebra, such as $\mathfrak{su}(2)$. We give some general results about the monodromies of such flat connections in the configuration spaces of points, and provide explicit examples of topological invariants of links (more generally, tangles) realized by the monodromy.   }

\date{}

\begin{document}
\maketitle
\section{Introduction}
The bulk-boundary correspondence of topological field theories(TQFTs) states that a $d$-dimensional bulk TQFT determines a $d-1$ dimensional field theory on its boundary by canceling the boundary anomaly, while the anomalous boundary theory can in turn reconstruct the bulk theory \cite{Callan1985,Witten1989,Elitzur1989,Moore1989,Wen1990,Muller2019,Gaiotto2021,debray2023longexactsequence}. A classical example of this correspondence is the relation between $3d$ Chern-Simons (CS) theory and the chiral $2d$ Wess-Zumino-Witten(WZW) model \cite{Witten1989}. In particular, within this correspondence, the expectation values of Wilson lines in the CS bulk can be derived from properties of WZW conformal blocks on the boundary.

Chern-Simons theory also describes the low-energy physics of anyon systems~\cite{71b20735d69d490e8f291b0760c40406}, where Wilson lines represent the worldlines of anyons. Rephrased in the language of anyon physics, the above correspondence implies that the fusion rules of anyons are determined by the operator product expansions (OPEs) in the WZW model, while their braiding statistics are encoded in the monodromy of conformal blocks. 

The $n$-point conformal blocks $\psi$ of WZW primary fields form a horizontal section of a vector bundle equipped with a flat connection given by the Knizhnik-Zamolodchikov~(KZ) equations \cite{KNIZHNIK198483}:
\begin{equation}
\label{eq:the KZ equation}
    \kappa\partial_i\psi=\sum_{i\neq j}\frac{\Omega_{ij}}{z_i-z_j}\psi
\end{equation}
where $i=1,\dots,n$ and $z_i$'s denote the insertion points of fields. The operators $\Omega_{ij}$ are constructed from tensor products of Lie algebra operators $J_i$ associated with a Lie algebra $\mathfrak{g}$, reflecting the $\hat{\mathfrak{g}}_\kappa$ Kac-Moody symmetry of the theory. When $z_i$ approaches $z_j$, the $\frac{1}{z_i-z_j}$ becomes singular; such first-order pole singularities are called regular singular points. From the representation-theoretic perspective, these singularities arise from highest weight representations of the affine algebra $\hat{\mathfrak{g}}_\kappa$.

The monodromies of these conformal blocks yields braid group representations. After taking a suitable trace over spin indices, one obtains quantum knot invariants~\cite{reshetikhin1990ribbon,kohno2002conformal}. These invariants are polynomials in $q=\exp(\frac{2\pi i}{\kappa})$. Expanding these polynomials in powers of $\frac{1}{\kappa}$, the coefficients themselves define knot invariants known as \textit{Vassiliev} invariants~\cite{inbook,bar1995vassiliev,birman1993knot}. The computation of such invariants is perturbative, meaning it can be carried out order by order and, in principle, can be obtained for arbitrary $n$-point functions to any desired order. In such a perturbative approach, one can also consider $\Omega_{ij}$ as generators of a formal algebra satisfying certain relations (required by the flatness condition). The result is then a \textit{universal} perturbative invariant, which can be obtained as \textit{Kontsevich integral} (see~\cite{chmutov2011introductionvassilievknotinvariants} for a textbook account).

A natural generalization of the KZ equations is obtained by allowing irregular singularities, which introduce higher-order poles of the form $\frac{1}{(z_i-z_j)^l}$ with integers $l>1$. Such equations were first introduced in \cite{Resh-KZ} and were then later studied from the vantage point of integrability in \cite{FMTV00,Jimbo_2008,Nagoya_2010}. In the context of WZW models, such irregular KZ equations were studied for the $H_3^+$ WZW model in \cite{Gaiotto:2013rk}. For $\frak{g}=\frak{su}(2)$, the representation-theoretic origin of irregular KZ equations has recently been studied in detail in \cite{gukov2025irregularkzequationskacmoody}, where irregular Kac-Moody representations are defined. However, the monodromy and fusion properties of the corresponding irregular fields have not yet been systematically investigated. 

One application of such generalized Kac-Moody modules is within the realm of class $S$ theories \cite{Gaiotto:2009we,Gaiotto:2009hg}. When compactifying on Gaiotto curves with irregular singular points, one obtains four-dimensional Argyres-Douglas (AD) theories \cite{Argyres:1995jj}, which are intrinsic superconformal field theories (SCFTs) without a Lagrangian description. Within the AGT correspondence \cite{Alday:2009aq}, AD theories correspond to Liouville conformal blocks in the presence of irregular (Whittaker) representations of the Virasoro algebra which were extensively studied in the past, see for example \cite{Awata:2009ur,Awata:2010bz,Nishinaka:2012kn,Rim:2012tf,Choi:2014qha}. Degenerate operators of Liouville theory on the curve then correspond to surface operators within the 4d AD SCFT and braiding of such operators is then governed by the irregular KZ connection \cite{gukov2025irregularkzequationskacmoody}. This then opens the path towards studying gauge theories obtained via irregular punctures from the perspective of Chern-Simons theory in the presence of irregular defects, see for example \cite{gaiotto2024sykschurdualitydoublescaled} for a recent application.

The goal of this paper is to study these properties and, in particular, to define a new class of link invariants associated with irregular singularities. We expect that the results will lead to novel mathematical structures such as generalized forms of fusion categories. From the perspective of the bulk-boundary correspondence, such developments may also shed light on the role of irregular singularities in the $3d$ anyon picture.

We find that for closures of braids that contain only one strand corresponding to an irregular singularity, the invariants turn out to be the same as in the case when all singularities are regular. However, this is no longer the case when there are two or more irregular singularities, or, instead of the closure of the braid, one considers a tangle produced by taking a certain limit of the braid.

The structure of the paper is as follows. In Section \ref{sec: Braids}, we review the basic mathematics of braids, links, and knots, and explain how their invariants arise from flat connections. In Section \ref{sec:KZ}, we introduce the KZ equation (both regular and irregular) and analyze its flatness condition, together with general methods for computing the monodromy (parallel transport). In Section \ref{sec:ex}, we apply these methods to specific examples and present novel link (and tangle) invariants. In the appendices, we include several lengthy computations, in particular the derivation of the associator.

\section{Braids, links and the flat connections}
\label{sec: Braids}

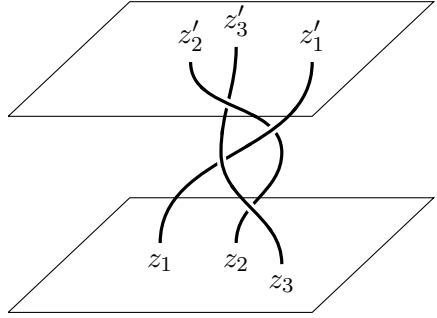
\begin{figure}[ht!]
\centering

\begin{tikzpicture}[scale=0.4,baseline=0]
\begin{knot}[flip crossing=1,flip crossing=3]
  \strand[very thick] (-0.5, 0)  node[below]{$z_2$} 
  to [out=up, in=down]  (1,3)
  to [out=up, in=down]  (-2,6) node[above]{$z_2'$};
\strand[very thick] (1, -0.7)   node[below]{$z_3$}
  to [out=up, in=down]  (-1,3)
  to [out=up, in=down]  (-0.5,6.5) node[above]{$z_3'$};
   \strand[very thick] (-3, 0)  node[below]{$z_1$}
  to [out=up, in=down]  (2,6)  node[above]{$z_1'$};
\end{knot}
\draw (-8,-2.3) -- (2,-2.3) -- (6,1.5) -- (-4,1.5) -- (-8,-2.3);
\draw (-8,4.2) -- (2,4.2) -- (6,8) -- (-4,8) -- (-8,4.2);
\end{tikzpicture}

\caption{A path from $(z_1,z_2,z_3)$ to $(z_1',z_2',z_3')$ in $Conf_3$, the configuration space of 3 points in $\mathbf{C}$. The vertical direction corresponds to the interval $I$ parametrizing the path, while the horizontal directions correspond to $\mathbf{C}$.}
\label{fig:conf-path}
\end{figure}

Let $Conf_n$ be the configuration space of $n$ points in $\mathbf{C}\cong \mathbf{R}^2$. Namely, it is the set of ordered $n$-tuples of distinct points in $\mathbf{C}$: $Conf_n:=\{(z_1,z_2,\ldots,z_n)|z_i\neq z_j,\;\forall i\neq j\}\subset \mathbf{C}^n$. A path in $Conf_n$, that is a continuous map from an interval $I=[0,1]$, provides an embedding of a disjoint union of $n$ copies of $I$ into $\mathbf{C}\times I$ such that its image intersects $\mathbb{C}\times \{t\}$ at exactly $n$ points for any $t\in I$. See Figure \ref{fig:conf-path} for an example.  We will assume that the path is smooth, which would imply the smoothness of the embedding of the intervals in $\mathbf{C}\times I$. The embedding can be understood as a special type of a \textit{tangle} in $\mathbf{C}\times I$, with the images of individual intervals being its \textit{strands}.

Homotopic paths correspond to isotopic (that is, ``topologically equivalent'') embeddings. As for paths in general topological spaces, a path from $(z_1,\ldots,z_n)\in Conf_n$ to $(z_1',\ldots,z_n')$ can be composed, i.e. concatenated\footnote{The composition in general will produce embedding which is not smooth at the place of the concatenation, unless one assumes vanishing of the derivative of the path near the endpoints of the interval. Nevertheless, the produced tangle can be always ``smoothed out''.}, with a path from $(z_1',\ldots,z_n')$ to $(z_1'',\ldots,z_n'')$ to get a path from $(z_1,\ldots,z_n)$ to $(z_1'',\ldots,z_n'')$. In the embedding picture illustrated in Figure \ref{fig:conf-path}, this corresponds to ``stacking'' two tangles on top of each other.

As usual, one can consider a special class of paths -- loops that start and end at the same fixed basepoint in $Conf_n$. One often chooses it to be $(0,1,2,\ldots,n-1)\in Conf_n$. That is, in the embedding picture, the strands start and end at the same points $0,1,\ldots,n-1$ on the real line inside $\mathbf{C}$. In this case, the embedding of the intervals provides a \textit{pure braid} with $n$ strands. See Figure \ref{fig:braids}~(b).

It is often also useful to consider the \textit{unordered} configuration space $Conf_n/S_n$ where $S_n$ is the symmetric group permuting the $n$ points. That is, it is the space of unordered $n$-tuples of distinct points. The embeddings of the intervals corresponding to the loops in this space are general braids. See Figure \ref{fig:braids} (a) for an example.

The fundamental groups of $Conf_n/S_n$ and $Conf_n$ are known as \textit{braid group} $Br(n)$ and \textit{pure braid group} $PBr(n)$ respectively. The elements in the groups correspond to isotopy classes of (pure) braids. The multiplication corresponds to the concatenation of the braids.

\begin{figure}[h]
  \centering
  \begin{subfigure}{0.45\textwidth}
    \centering
    \begin{tikzpicture}[
braid/.cd,
]
\pic[braid/crossing height=1.34cm] at (0,0){braid={s_1 s_2^{-1} s_1}};
\end{tikzpicture}
  \caption{A general braid}
  \end{subfigure}
  \hfill
  \begin{subfigure}{0.45\textwidth}
    \centering
    \begin{tikzpicture}[
braid/.cd,]
\pic at (0,0){braid={s_1 s_1 s_2 s_2}};
\end{tikzpicture}
    \caption{A pure braid}
  \end{subfigure}
  \caption{Braids}
  \label{fig:braids}
\end{figure}

The action of the symmetric group $S_n$ is natural on the vector bundles and connections introduced in the following sections. Although our discussion will, by default, assume the pure braid setting, the extension to general braids is straightforward. Accordingly, we will not distinguish between “pure braids” and “braids” unless such a distinction is necessary; the precise meaning should be clear from the context.

\subsection{The monodromy representation of the braid group}
We would like to consider braid invariants, that is, quantities associated with a braid that are invariant under smooth deformations of the strands. In more formal terms, a braid invariant is a map from the set of isotopy classes of braids to some other set $X$. If $X$ is a group and the invariant is compatible with braid multiplication, then it defines a representation of the braid group. Furthermore, if $X=GL(V)$ for some vector space $V$, then this map, which we denote by $\rho$, is a linear representation of the braid group.

As mentioned above, the braid group can be realized as the fundamental group of a configuration space. This realization provides a geometric method for constructing braid group representations, namely via monodromy representations. Consider a vector bundle $E \rightarrow Conf_n$ equipped with a flat connection $\nabla$, and denote its fiber by $V$.

In a local trivialization, the connection takes the form
\begin{equation}
\nabla = d + \omega,
\end{equation}
where $d$ is the exterior derivative and $\omega$ is a $\mathfrak{gl}(V)$-valued 1-form. The flatness condition for the connection is given locally by $d\omega+\omega\wedge\omega=0.$

Parallel transport along a path $\gamma:I\rightarrow Conf_n$ of a vector is defined by the equation $\nabla_{\dot\gamma} v = 0$. Solving this differential equation yields the \textit{holonomy} 
\begin{equation}
\label{eq:monodromy form}
    M(\gamma)=P\exp(-\int_\gamma\omega),
\end{equation} 
where $P$ denotes the path ordering.
That is, parallel transport along a path $\gamma$ maps a vector $v \in V$ to $P\exp(-\int_\gamma\omega)\cdot v$. Since the connection is flat, the result depends only on the homotopy class of the path $\gamma$, and thus the holonomy can be refered to as \textit{monodromy}. Consequently, parallel transport along loops defines a group homomorphism
\begin{equation}
\rho_{\nabla} : \pi_1(Conf_n) \rightarrow GL(V).
\end{equation}
This homomorphism is precisely the monodromy representation of the braid group.

\subsection{From braids to links and tangles}
A knot is an embedding $K:S^1\rightarrow M$ of a circle $S^1$ into a $3$-manifold $M$. An embedding of multiple circles is referred to as a link, with a knot thus being a particular type of link. 

\begin{figure}[ht!]
\centering
\begin{subfigure}{0.3\textwidth}
\centering
\begin{tikzpicture}[scale=0.4,baseline=0]
\begin{knot}[flip crossing=1]
  \strand (-1, 0)  
  to [out=up, in=down]  (1,3)
  to [out=up, in=down]  (-1,6);
\strand (1, 0)  
  to [out=up, in=down]  (-1,3)
  to [out=up, in=down]  (1,6);
\end{knot}
\end{tikzpicture}
\caption{}
\end{subfigure}
\begin{subfigure}{0.5\textwidth}  
\centering
\begin{tikzpicture}[scale=0.4,baseline=0]
\begin{knot}[flip crossing=1]
  \strand (-1, 0)  
  to [out=up, in=down]  (1,3)
  to [out=up, in=down]  (-1,6)
  to [out=up, in=right] (-2,7.5)
  to [out=left, in=up] (-4,3)
  to [out=down, in=left] (-2,-1.5)
  to [out=right,in=down] (-1,0);
 
\strand (1, 0)  
  to [out=up, in=down]  (-1,3)
  to [out=up, in=down]  (1,6)
  to [out=up, in=right] (-2,8)
  to [out=left, in=up] (-5,3)
  to [out=down, in=left] (-2,-2)
  to [out=right,in=down] (1,0);
\end{knot}
\draw[very thick] (1.5, 0)  
  to [out=up, in=down]  (1.5,6)
  to [out=up, in=right] (-2,8.5)
  to [out=left, in=up] (-5.5,3)
  to [out=down, in=left] (-2,-2.5)
  to [out=right,in=down] (1.5,0);

\draw[very thick] (-1.5, -0.5)  
  to [out=up, in=down]  (-1.5,6.5);

\draw[very thick] (-1.5, 6)
  to [out=-130, in=up] (-3,3)
  to [out=down, in=130]  (-1.5,0);

\end{tikzpicture}
\caption{}
\end{subfigure}
\caption{A braid and its closure in a solid torus $D^2\times S^1$.}
\label{fig:braid-closure-solid-torus}
\end{figure}
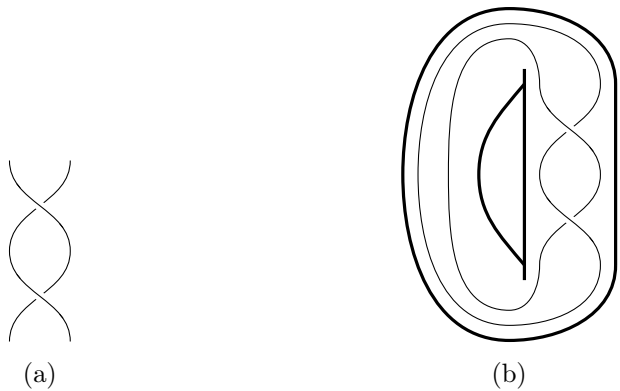

There is a natural, although by no means unique, way to obtain links and their invariants via closure of braids. Consider a braid embedded in the subspace $D^2\times I\subset \mathbf{R}^3$. Here $I$ is the time direction and all strands begin on one copy of the disk $D^2$ and end on another copy of $D^2$. Identifying these two copies of $D^2$ produces a solid torus $D^2\times S^1$. If we further identify the beginning and end points of each strand, we obtain a link living in the solid torus, see Figure \ref{fig:braid-closure-solid-torus}.  


As discussed above, to each braid in $D^2\times I$, we can associate a map (the monodromy representation) taking values in $GL(V)$ (or, instead, a formal algebra with some generators and relations). 
Considering, up to isotopy, the link obtained by the closure of a braid is equivalent to forgetting the basepoint of the loop in $Conf_n$ and remembering only the homotopy class of the map $S^1\rightarrow Conf_n$. For such a homotopy class, the monodromy of the KZ connection is defined only up to conjugation. The trace of the monodromy is invariant under conjugation and thus naturally yields a link invariant. 
In the following sections, we will often convert braids to links using this particular closure procedure.

\begin{figure}[ht!]
\centering
\begin{subfigure}{0.3\textwidth}
\centering
\begin{tikzpicture}[scale=0.4,baseline=0]
\begin{knot}[flip crossing=1,flip crossing=3]
  \strand (-0.5, 0)  
  to [out=up, in=down]  (1,3)
  to [out=up, in=down]  (-0.5,6);
\strand (0.5, 0)  
  to [out=up, in=down]  (-1,3)
  to [out=up, in=down]  (0.5,6);
   \strand (-3, 0)  node[left]{$-\infty\leftarrow$}
  to [out=up, in=down]  (0,3)
  to [out=up, in=down]  (3,6)  node[right]{$\rightarrow+\infty$};
\end{knot}
\end{tikzpicture}
\caption{}
\end{subfigure}
\begin{subfigure}{0.5\textwidth}  
\centering
\begin{tikzpicture}[scale=0.4,baseline=0]
\begin{knot}[flip crossing=1,flip crossing=3]
  \strand (-0.5, 0)  
  to [out=up, in=down]  (1,3)
  to [out=up, in=down]  (-0.5,5.5)
  to [out=up, in=right] (-2,7.5)
  to [out=left, in=up] (-4,3)
  to [out=down, in=left] (-2,-1.5)
  to [out=right,in=down] (-0.5,0);
\strand (0.5, 0)  
  to [out=up, in=down]  (-1,3)
  to [out=up, in=down]  (0.5,5.5)
  to [out=up, in=right] (-2,8)
  to [out=left, in=up] (-5,3)
  to [out=down, in=left] (-2,-2)
  to [out=right,in=down] (0.5,0);
   \strand (-1.5,5.5)
   to[out=right,in=down]   (-1,6.5)   
   to [out=up,in=right] (-2,7)
   to[out=left,in=up ](-3.5,3)
   to[out=down,in=left] (-2,-1)
   to[out=right, in=down] (-1, 0)  
  to [out=up, in=down]  (0.5,3)
  to [out=up, in=left]  (1.5,5.5);
\end{knot}
\draw[very thick] (1.5, 0)  
  to [out=up, in=down]  (1.5,6)
  to [out=up, in=right] (-2,8.5)
  to [out=left, in=up] (-5.5,3)
  to [out=down, in=left] (-2,-2.5)
  to [out=right,in=down] (1.5,0);

\draw[very thick] (-1.5, -0.5)  
  to [out=up, in=down]  (-1.5,6.5);

\draw[very thick] (-1.5, 6)
  to [out=-130, in=up] (-3,3)
  to [out=down, in=130]  (-1.5,0);

\end{tikzpicture}
\caption{}
\end{subfigure}
\caption{A tangle in $\mathbf{C}\times I$ and a closure of its limit in the solid torus $D^2\times S^1$.}
\label{fig:tangle-closure-solid-torus}
\end{figure}
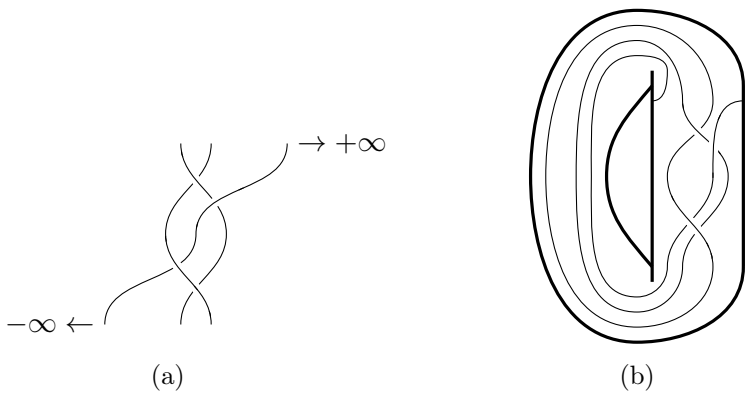

We will also consider a setting where one takes a limit in which the initial and terminal positions of some of the strands are taken to infinity in $\mathbf{C}$ simultaneously, in different directions (e.g. opposite). This limit will be taken simultaneously with the limit in which some parameter in the connection is taken to zero, with a certain scaling relation imposed. After identifying the endpoints of $I$ this produces a \textit{tangle} in $\mathbf{C}\times S^1\cong D^2\times S^1$ which is an embedding of a disjoint union of copies of $S^1$ and intervals, such that the images of the endpoints of the intervals are on the boundary of the solid torus (see Figure \ref{fig:tangle-closure-solid-torus}). After such a closure, one is allowed to freely move from top to bottom, and vice versa, braids of strands that are not ending at the boundary. This would again result into a change of the parallel transport $P
\exp -\int \omega$ by a conjugation. Taking the trace gets rid of such an ambiguity.

\section{The KZ equation}
\label{sec:KZ}
In coordinates $(z_1,\ldots,z_n)$ of the space $Conf_n$, the KZ connection formally takes the form:
\begin{equation}
\label{eq:regular connection}
    \nabla=d-\sum_{1\leq i<j\leq n}\Omega_{ij}\frac{dz_i-dz_j}{z_j-z_i}
\end{equation}
where the operators $\Omega_{ij}$ takes values in $\mathfrak{gl}(V)$. It realizes the KZ equation (\ref{eq:the KZ equation}) with $\kappa=1$ as the equation on a flat section $\psi$. Using the notation from the previous section, we may write the connection 1-form $\omega=-\sum_{1\leq i<j\leq n}\Omega_{ij} d \log(z_i-z_j)$, from which it follows that $d\omega=0$.

We deliberately refrain from specifying the vector space $V$ or the precise action of the operators $\Omega_{ij}$. Our goal is instead to derive general conditions on the $\Omega_{ij}$ imposed by the flatness of the connection. A convenient way to do this is to work component-wise. Define
\begin{equation}
    \omega_i=-\sum_{k\neq i}\frac{\Omega_{ik}}{z_i-z_k}
\end{equation}
where we set $\Omega_{ki}=\Omega_{ik}$ for $k>i$. One readily verifies that $\omega=\sum_i\omega_idz_i$. The condition $d\omega=0$ then translates into $\partial_i\omega_j-\partial_j\omega_i=0$. 

The flatness condition reduces to the requirement:
\begin{equation}
    [\omega_i,\omega_j]=\sum_{k\neq i}\sum_{l\neq j}[\frac{\Omega_{ik}}{z_i-z_k},\frac{\Omega_{jl}}{z_j-z_l}]=0.
\end{equation}
When the indices $i,j,k,l$ are all distinct, the functions $\frac{1}{(z_i-z_k)(z_j-z_l)}$ are linearly independent from those involving other combinations of labels. Consequently, the coefficient of these terms, namely the commutators $[\Omega_{ik},\Omega_{jl}]$ must vanish individually. 

However, when considering terms like $[\Omega_{ik},\Omega_{jk}]$, one must take into account relations among the corresponding rational functions, for example
\begin{equation}
    -\frac{1}{(z_i-z_k)(z_j-z_k)}=\frac{1}{(z_i-z_j)(z_j-z_k)}+\frac{1}{(z_j-z_i)(z_i-z_k)}.
\end{equation}
As a result, the functionally independent contributions are $\frac{[\Omega_{ij}+\Omega_{ik},\Omega_{jk}]}{(z_i-z_j)((z_j-z_k)}$ and $\frac{[\Omega_{ij}+\Omega_{jk},\Omega_{ik}]}{(z_j-z_i)(z_i-z_k)}$.
From these observations, we conclude that the operators $\Omega_{ij}$ must satisfy the following conditions
\begin{equation}
\label{eq:condition}
\begin{gathered}
[\Omega_{ij}+\Omega_{jk},\Omega_{ik}]=0, \quad i,j,k\quad \text{distinct,}\\
      [\Omega_{ij},\Omega_{kl}]=0,\quad i,j,k,l\quad \text{distinct.}
\end{gathered}
\end{equation}
These relations are known as the infinitesimal pure braid relations.

\subsection{The space of chord diagrams and the associator}
Without further specification of the vector space $V$ and the action of the operators $\Omega_{ij}$, little can be said about the resulting holonomies. However, there exists a universal choice of $V$ for which minimal additional input is required: namely, the case in which  $V$ is taken to be the space generated by the $\Omega_{ij}$ themselves. 

We now define this space more precisely. For an integer $n>1$, let $\mathcal{A}^h(n)$ be the algebra generated by symbols $\Omega_{ij}, 1\leq i<j\leq n$, subject to the relations \eqref{eq:condition}. Equivalently, $\mathcal{A}^h(n)$ consists of all formal polynomials in the generators $\Omega_{ij}$ modulo these relations. This algebra is known as the algebra of horizontal chord diagrams on $n$ strands. 

Each generator $\Omega_{ij}$ admits a diagrammatic representation as a horizontal chord connecting the $i$-th and the $j$-th strands:
\begin{equation}
    \Omega_{ij}=\;
    \begin{tikzpicture}[scale=0.5,baseline=(current  bounding  box.center)]
        \draw (2,-2) node[below] {$i$} -- (2,2)  ;
        \draw (5,-2) node[below] {$j$} -- (5,2);
        \begin{scope}[red,decoration={markings,mark=at position 0 with {\fill[red] circle (0.05);},mark=at position 1 with {\fill[red] circle (0.05);}}]
            \draw[postaction=decorate] (2,0) -- (5,0);
        \end{scope}
        
    \end{tikzpicture}
    .
\end{equation}
The multiplication in $\mathcal{A}^h(n)$ is represented by vertical juxtaposition, that is, by placing one diagram below another. The multiplicative unit is denoted by $1_n$, whose diagram consists of $n$ straight, unconnected strands. The relations (\ref{eq:condition}) then have the following graphical represenation:
\begin{equation}
     \begin{tikzpicture}[scale=0.5,baseline=(current  bounding  box.center)]
        \draw (0,-2) node[below] {$i$} -- (0,2)  ;
        \draw (1.5,-2) node[below] {$j$} -- (1.5,2);
        \draw (3,-2) node[below] {$k$} -- (3,2);
        \begin{scope}[red,decoration={markings,mark=at position 0 with {\fill[red] circle (0.05);},mark=at position 1 with {\fill[red] circle (0.05);}}]
            \draw[postaction=decorate] (0,0.5) -- (1.5,0.5);
            \draw[postaction=decorate] (0,-0.5) -- (3,-0.5);
        \end{scope}  
    \end{tikzpicture}
    +
    \begin{tikzpicture}[scale=0.5,baseline=(current  bounding  box.center)]
        \draw (0,-2) node[below] {$i$} -- (0,2)  ;
        \draw (1.5,-2) node[below] {$j$} -- (1.5,2);
        \draw (3,-2) node[below] {$k$} -- (3,2);
        \begin{scope}[red,decoration={markings,mark=at position 0 with {\fill[red] circle (0.05);},mark=at position 1 with {\fill[red] circle (0.05);}}]
            \draw[postaction=decorate] (1.5,0.5) -- (3,0.5);
            \draw[postaction=decorate] (0,-0.5) -- (3,-0.5);
        \end{scope}  
    \end{tikzpicture}
    -
    \begin{tikzpicture}[scale=0.5,baseline=(current  bounding  box.center)]
        \draw (0,-2) node[below] {$i$} -- (0,2)  ;
        \draw (1.5,-2) node[below] {$j$} -- (1.5,2);
        \draw (3,-2) node[below] {$k$} -- (3,2);
        \begin{scope}[red,decoration={markings,mark=at position 0 with {\fill[red] circle (0.05);},mark=at position 1 with {\fill[red] circle (0.05);}}]
            \draw[postaction=decorate] (0,-0.5) -- (1.5,-0.5);
            \draw[postaction=decorate] (0,0.5) -- (3,0.5);
        \end{scope}  
    \end{tikzpicture}
    -
    \begin{tikzpicture}[scale=0.5,baseline=(current  bounding  box.center)]
        \draw (0,-2) node[below] {$i$} -- (0,2)  ;
        \draw (1.5,-2) node[below] {$j$} -- (1.5,2);
        \draw (3,-2) node[below] {$k$} -- (3,2);
        \begin{scope}[red,decoration={markings,mark=at position 0 with {\fill[red] circle (0.05);},mark=at position 1 with {\fill[red] circle (0.05);}}]
            \draw[postaction=decorate] (1.5,-0.5) -- (3,-0.5);
            \draw[postaction=decorate] (0,0.5) -- (3,0.5);
        \end{scope}  
    \end{tikzpicture}
    =0
\end{equation}
known as 4-term relation on chord diagrams. The chords connecting pairwise distinct strands can be slid past each other:
\begin{equation}
     \begin{tikzpicture}[scale=0.5,baseline=(current  bounding  box.center)]
        \draw (0,-2) node[below] {$i$} -- (0,2)  ;
        \draw (1.5,-2) node[below] {$j$} -- (1.5,2);
        \draw (3,-2) node[below] {$k$} -- (3,2);
         \draw (4.5,-2) node[below] {$l$} -- (4.5,2);
        \begin{scope}[red,decoration={markings,mark=at position 0 with {\fill[red] circle (0.05);},mark=at position 1 with {\fill[red] circle (0.05);}}]
            \draw[postaction=decorate] (0,0.5) -- (1.5,0.5);
            \draw[postaction=decorate] (3,-0.5) -- (4.5,-0.5);
        \end{scope}  
    \end{tikzpicture}
    \; = \;
    \begin{tikzpicture}[scale=0.5,baseline=(current  bounding  box.center)]
        \draw (0,-2) node[below] {$i$} -- (0,2)  ;
        \draw (1.5,-2) node[below] {$j$} -- (1.5,2);
        \draw (3,-2) node[below] {$k$} -- (3,2);
         \draw (4.5,-2) node[below] {$l$} -- (4.5,2);
        \begin{scope}[red,decoration={markings,mark=at position 0 with {\fill[red] circle (0.05);},mark=at position 1 with {\fill[red] circle (0.05);}}]
            \draw[postaction=decorate] (3,0.5) -- (4.5,0.5);
            \draw[postaction=decorate] (0,-0.5) -- (1.5,-0.5);
        \end{scope}  
    \end{tikzpicture}
    .
\end{equation} 

Some low-$n$ cases can be worked out explicitly. The algebra $\mathcal{A}^h(2)$ is simply the free commutative algebra generated by a single element $\Omega_{12}$. The algebra $\mathcal{A}^h(3)$ decomposes as the direct product of the free~(non-commutative) algebra generated by $\Omega_{12}$ and $\Omega_{23}$, and the free commutative algebra generated by the central element:
\begin{equation}
    u=\Omega_{12}+\Omega_{23}+\Omega_{13}.
\end{equation}

Within the framework of $\mathcal{A}^h(n)$, the solutions of the KZ equations (flat sections), the monodromy and the associator (to be introduced below) are to be considered as elements in a completion of $\mathcal{A}^h(n)$: formal power series in the non-commutative variables $\Omega_{ij}$ modulo the relation. The terms in the power series have natural grading by the total degree in the generators $\Omega_{ij}$. Consequently, computations are carried out order by order with respect to this grading. Throughout this section, all functions of the generators $\Omega_{ij}$ are to be understood as formal power series. For example, $\exp(\Omega_{ij})=1+\Omega_{ij}+\frac12\Omega_{ij}^2+\ldots$

With this understanding we are now ready to make the formal expression for the monodromy in \eqref{eq:monodromy form} more concrete by working within the algebra $\mathcal{A}^h(n)$. This relies on the existence of distinguished solutions $\psi_{ij}$ with the asymptotic behavior
\begin{equation}
\label{eq:asymptotics}
    \psi_{ij}\sim (z_i-z_j)^{\Omega_{ij}}.
\end{equation}

Along a small loop encircling the divisor $z_i=z_j$, the monodromy action on $\psi_{ij}$ takes the particular simple form $\exp(2\pi i\Omega_{ij})$. Since the braid group is generated by such elementary loops, any monodromy representation can be decomposed into products of these basic elements. 

What remains to be understood is the transformation relating solutions $\psi_{ij}$ corresponding to different pairs of indices $i,j$. Because all $\psi_{ij}$ are solutions of the same linear differential equation, they differ by an element of $GL(V)$. These transition matrices, which we denote by $\Psi$, are called the associator. Formally the associator relating $\psi_{ij}$ and $\psi_{kl}$ can be written as
\begin{equation}
\label{eq:def of asso}
    \Psi=\psi_{ij}^{-1}\psi_{kl}.
\end{equation}
Any monodromy can therefore be expressed as a combination of associators $\Psi$ and the local monodromy factors $\exp(2\pi i\Omega_{ij})$. 

Due to the topological nature of the construction, the form of $\Psi$ is universal: it is the same in the higher-point case as in the lower-point case. In particular, the three-point case already suffices to determine $\Psi$. Since explicit computation is somewhat involved, we defer it to the appendix, while freely using the resulting expression in the main text.

Finally, let us comment on restrictions of the configuration space coordinates $z_i$. One may “freeze” any coordinate $z_j$ to a fixed value and consider the resulting system in the remaining variables; this effectively removes the equation involving $\partial_{z_j}$. In the three-dimensional picture, this corresponds to keeping the strand at $z_j$ fixed while allowing the others to move. In this restricted setting, one obtains only pure braid monodromies around $z_j$, but not braidings involving $z_j$ itself. A particularly convenient case is the reduction to an ordinary differential equation, where a single variable is allowed to vary while all others are fixed. In Section \ref{sec:ex}, we will work in this simplified setting.

\subsection{Matrix forms and WZW realizations}
There is a natural realization of $V$ and $\Omega_{ij}$ as follows. One may take $V$ to be a complex vector space $\mathbf{C}^m$ for some $m$, and $\Omega_{ij}$ to be $m\times m$ matrices satisfying the flatness condition \eqref{eq:condition}. In this way, the KZ equation is reduced to a system of matrix differential equations. The functions $\psi_{ij}$ appearing in \eqref{eq:asymptotics} are then genuine matrix-valued functions, and the inverse in \eqref{eq:def of asso} is understood as the usual matrix inverse.

In what follows, when we refer to an \textit{exact} solution of the KZ equation, we mean a solution in this matrix framework.

An important class of such matrix realizations arises from the WZW model. In this setting, one begins by choosing a simple Lie algebra, for example $\mathfrak{g}=\mathfrak{su}(2)$, and assigning to each strand $i$ a highest-weight representation $V_i$ of this Lie algebra. One then sets $V=\bigotimes_i V_i$ and defines $\Omega_{ij}=\sum_\mu J_{i}^\mu\otimes J_j^\mu$, where $J_k^\mu$ are the generators of the Lie algebra acting on the representation $V_k$. The flatness conditions (\ref{eq:condition}) follow directly from the commutation relations satisfied by the generators $J_j^\mu$.

Let us write down the matrices explicitly in the simplest Lie algebra example. We take the Lie algebra to be $\mathfrak{su}(2)$, whose generators are the angular momentum operators $J^z,J^x,J^y$. Consider the $2$-point case and take $V_1,V_2$ to be the $2$-dimensional spin-$\frac12$ representation, with basis vectors denoted by $\uparrow$ and $\downarrow$. In this basis ${\uparrow,\downarrow}$ the generators read
\begin{equation}
J^z= \begin{pmatrix}
\frac12&0\\
0&-\frac12
\end{pmatrix},J^x=\begin{pmatrix}
0&\frac12\\
\frac12&0
\end{pmatrix},J^y=\begin{pmatrix}
0&\frac{i}2\\
-\frac{i}{2}&0
\end{pmatrix}.
\end{equation}
We then take $V=V_1\otimes V_2$, whose basis we choose to be ${\uparrow\uparrow,\uparrow\downarrow,\downarrow\uparrow,\downarrow\downarrow}$. Then
\begin{equation}
\label{eq: matrix rep}
\Omega_{12}\doteq\begin{pmatrix}
\frac12&&&\\
&-\frac12&1&\\
&1&-\frac12&\\
&&&\frac12
\end{pmatrix}
.
\end{equation}
In general, the size of the matrix $\Omega_{ij}$ grows as $2^n$ for the $n$-point equation.

In the standard WZW model formulation, the KZ equation is often written as in (\ref{eq:the KZ equation} with non-trivial $\kappa$ constant related to the Kac–Moody level. Introducing $\kappa$ does not alter the essential structure of the analysis, but amounts to a rescaling of $\Omega_{ij}$. One may therefore keep track of the dependence of solutions on $\kappa$ explicitly. Expanding the exact solution in powers of $\frac1\kappa$ reproduces the formal order-by-order computation introduced in Section~3.1. We illustrate this correspondence with several examples in the appendix. In what follows, we will freely switch between including $\kappa$ explicitly and omitting it, as this does not affect the underlying arguments.

\subsection{Irregular singularities}
The singularities of \eqref{eq:regular connection} occur along the diagonals $z_i = z_j$ and are of the form $\frac{1}{z_i-z_j}$, which are first-order poles. We refer to such singularities as regular singularities. In contrast, irregular singularities are higher-order poles; in the present setting, they are terms of the form $\frac{1}{(z_i - z_j)^{l+1}}$ with integer $l \geq 1$. The integer $l$ is called the rank of the irregular singularity.

In general, an irregular KZ connection takes the form
\begin{equation}
\label{eq:irregular connection}
\nabla = \nabla_{\mathrm{reg}}-
\sum_{\substack{1 \le i < j \le n \\ 1\leq l \leq l_{ij} }}
\Omega^{(l)}_{ij}\frac{d(z_i - z_j)}{(z_i - z_j)^{l+1}}-
\sum_{\substack{1\leq i\leq n\\ 1\leq l\leq l_{i\infty}}} H_i^{(l)} z_i^{l-1} dz_i 
\end{equation}
where $\nabla_\text{reg}$ is the connection of the form (\ref{eq:regular connection}). The terms $\sum_{i,l} H_i^{(l)} z_i^{l-1} dz_i$ represent irregular singularities at $z=\infty$. 

The flatness condition and the chord diagram algebra can be derived similarly. For simplicity, we focus on the case of a rank-1 irregular singularity at $z=\infty$. In this case, the connection $\omega$ takes the form
\begin{equation}
\label{eq:irregular connection}
\omega = \omega_i  dz_i
= -\left( H_i + \sum_{j \neq i} \frac{\Omega_{ij}}{z_i - z_j} \right) dz_i .
\end{equation}
One can verify that $d\omega=0$. Equivalently $\partial_j\omega_i-\partial_i\omega_j=0$. The flatness condition $\omega \wedge \omega = 0$ imposes, in addition to \eqref{eq:condition}, the following constraints:
\begin{equation}
\label{eq:flatness}
\begin{gathered}
[H_i, H_j] = 0 , \qquad \forall i,j , \\
[H_i + H_j, \Omega_{ij}] = 0,\\
[H_i, \Omega_{kl}] = 0 , \qquad i,k,l \ \text{distinct}. \\
\end{gathered}
\end{equation}

The algebra of the vertical chord diagrams can be extended to include the generators $H_i$, corresponding to dots supported on single strands:
 \begin{equation}
    H_i=\;
    \begin{tikzpicture}[scale=0.5,baseline=(current  bounding  box.center)]
        \draw (0,-2) node[below] {$i$} -- (0,2)  ;
        \fill[OliveGreen] (0,0) circle (0.1);
      \end{tikzpicture}.
\end{equation}
They satisfy the following relations:
\begin{equation}
    \begin{tikzpicture}[scale=0.5,baseline=(current  bounding  box.center)]
        \draw (0,-2) node[below] {$i$} -- (0,2)  ;
        \draw (1.5,-2) node[below] {$j$} -- (1.5,2)  ;
        \fill[OliveGreen] (0,0.5) circle (0.1);
        \fill[OliveGreen] (1.5,-0.5) circle (0.1);
      \end{tikzpicture}
      =
      \begin{tikzpicture}[scale=0.5,baseline=(current  bounding  box.center)]
        \draw (0,-2) node[below] {$i$} -- (0,2)  ;
        \draw (1.5,-2) node[below] {$j$} -- (1.5,2)  ;
        \fill[OliveGreen] (1.5,0.5) circle (0.1);
        \fill[OliveGreen] (0,-0.5) circle (0.1);
      \end{tikzpicture}
      ,
\end{equation}
\begin{equation}
     \begin{tikzpicture}[scale=0.5,baseline=(current  bounding  box.center)]
         \draw (1.5,-2) node[below] {$i$} -- (1.5,2);
        \draw (3,-2) node[below] {$k$} -- (3,2);
         \draw (4.5,-2) node[below] {$l$} -- (4.5,2);
        \begin{scope}[red,decoration={markings,mark=at position 0 with {\fill[red] circle (0.05);},mark=at position 1 with {\fill[red] circle (0.05);}}]
              \draw[postaction=decorate] (3,-0.5) -- (4.5,-0.5);
        \end{scope} 
        \fill[OliveGreen] (1.5,0.5) circle (0.1);
    \end{tikzpicture}
    \; = \;
    \begin{tikzpicture}[scale=0.5,baseline=(current  bounding  box.center)]
         \draw (1.5,-2) node[below] {$i$} -- (1.5,2);
        \draw (3,-2) node[below] {$k$} -- (3,2);
         \draw (4.5,-2) node[below] {$l$} -- (4.5,2);
        \begin{scope}[red,decoration={markings,mark=at position 0 with {\fill[red] circle (0.05);},mark=at position 1 with {\fill[red] circle (0.05);}}]
            \draw[postaction=decorate] (3,0.5) -- (4.5,0.5);
        \end{scope}  
        \fill[OliveGreen] (1.5,-0.5) circle (0.1);
    \end{tikzpicture}
    ,
\end{equation}
\begin{equation}
    \begin{tikzpicture}[scale=0.5,baseline=(current  bounding  box.center)]
        \draw (0,-2) node[below] {$i$} -- (0,2)  ;
        \draw (1.5,-2) node[below] {$j$} -- (1.5,2)  ;
        \fill[OliveGreen] (0,0.5) circle (0.1);
        \begin{scope}[red,decoration={markings,mark=at position 0 with {\fill[red] circle (0.05);},mark=at position 1 with {\fill[red] circle (0.05);}}]
              \draw[postaction=decorate] (0,-0.5) -- (1.5,-0.5);
        \end{scope} 
      \end{tikzpicture}
      +
      \begin{tikzpicture}[scale=0.5,baseline=(current  bounding  box.center)]
        \draw (0,-2) node[below] {$i$} -- (0,2)  ;
        \draw (1.5,-2) node[below] {$j$} -- (1.5,2)  ;
        \fill[OliveGreen] (1.5,0.5) circle (0.1);
        \begin{scope}[red,decoration={markings,mark=at position 0 with {\fill[red] circle (0.05);},mark=at position 1 with {\fill[red] circle (0.05);}}]
              \draw[postaction=decorate] (0,-0.5) -- (1.5,-0.5);
        \end{scope} 
      \end{tikzpicture}
      =
      \begin{tikzpicture}[scale=0.5,baseline=(current  bounding  box.center)]
        \draw (0,-2) node[below] {$i$} -- (0,2)  ;
        \draw (1.5,-2) node[below] {$j$} -- (1.5,2)  ;
        \fill[OliveGreen] (0,-0.5) circle (0.1);
        \begin{scope}[red,decoration={markings,mark=at position 0 with {\fill[red] circle (0.05);},mark=at position 1 with {\fill[red] circle (0.05);}}]
              \draw[postaction=decorate] (0,0.5) -- (1.5,0.5);
        \end{scope} 
      \end{tikzpicture}
      +
      \begin{tikzpicture}[scale=0.5,baseline=(current  bounding  box.center)]
        \draw (0,-2) node[below] {$i$} -- (0,2)  ;
        \draw (1.5,-2) node[below] {$j$} -- (1.5,2)  ;
        \fill[OliveGreen] (1.5,-0.5) circle (0.1);
        \begin{scope}[red,decoration={markings,mark=at position 0 with {\fill[red] circle (0.05);},mark=at position 1 with {\fill[red] circle (0.05);}}]
              \draw[postaction=decorate] (0,0.5) -- (1.5,0.5);
        \end{scope} 
      \end{tikzpicture}
            .
\end{equation}

As in the regular case, any matrices satisfying the flatness condition \eqref{eq:flatness} can be used as valid inputs for explicit computations.

In the conformal field theory setting, the matrix realization of $H_i$ arises from irregular representations of the Kac–Moody algebra \cite{gukov2025irregularkzequationskacmoody}. In this framework, we identify $H_i$ with $J^z_i$. For example, considering again the $2$-point case, in the basis ${\uparrow\uparrow,\uparrow\downarrow,\downarrow\uparrow,\downarrow\downarrow}$, the operator $H_2$ takes the form
\begin{equation}
\label{eq:irre matrix rep}
H_2\doteq\begin{pmatrix}
\frac12&&&\\
&-\frac12&&\\
&&\frac12&\\
&&&-\frac12
\end{pmatrix}.
\end{equation}

The existence of solutions of the form \eqref{eq:asymptotics} near regular singularities continues to hold in the generalized setting. As in the purely regular case, the monodromy can be decomposed into local exponents and the associator $\Psi$. The latter receives contributions from the irregular corrections $H_i$, whereas the local exponents $\exp(2\pi i\Omega_{ij})$ remain unaffected. See Appendix \ref{A:associator} for an explicit expression for $\Psi$.

In contrast, the monodromy around irregular singularities requires separate treatment.
\subsection{The Stokes phenomenon}
Let us consider the special case of an irregular KZ equation where we are looking at the differential equation for a regular point $z_i$ (where $i \neq 1$) in the presence of an irregular singularity at $z_1$. The regular point $z_i$ can be thought of arising from the insertion of a degenerate \textit{probe} operator, and the ``back-reaction'' of the irregular pole onto the regular sites gives rise to the corresponding KZ equation:
\begin{equation}
    \kappa \partial_i \psi = \left(\sum_{j \neq i} \frac{\Omega_{ij}}{z_i - z_j} + \sum_{l=1}^r \frac{\Omega_{1,i}^{(l)}}{(z_1 - z_i)^{l+1}}\right)\psi.
\end{equation}
Near $z_1$, the solution $\psi$ can be represented formally. However, because of the $1/(z_1 - z_i)^{l+1}$ terms, the formal series $\hat \psi$ typically has a zero radius of convergence. To get an actual holomorphic solution, you must apply \textit{Borel resummation} along a specific direction in the complex plane. The complex plane around $z_1$ is divided into $2r$ sectors (Stokes sectors). In each sector $S_k$, there exists a unique holomorphic solution $\psi_k$ that is asymptotic to the formal series $\hat \psi$. 

The Stokes phenomenon arises from the observation that $\psi_k \neq \psi_{k+1}$ even though they are asymptotic to the same formal series $\hat \psi$. When one crosses a \textit{Stokes line}, the subdominant part of the solution ``jumps''. The relationship is given by the Stokes matrix $S_k$:
\begin{equation}
    \psi_{k+1} = \psi_k \cdot S_k.
\end{equation}
The matrices $S_k$ are unipotent, meaning they only modify the exponentially small terms that the formal series $\hat \psi$ cannot see.

In our specific example, the leading behavior of the solution as $z_i \rightarrow z_1$ is governed by an exponential of a polynomial of degree $r$:
\begin{equation}
    \psi \sim \exp\left(\sum_{l=1}^r \frac{\Lambda_l}{(z_1 - z_i)^l}\right) (z_1 - z_i)^C (\mathrm{const}+\ldots),
\end{equation}
where $\Lambda_l$ are related to the eigenvalues of the irregular operators $\Omega^{(l)}_{1,i}$. The Stokes directions are the rays where the real part of the difference between these exponential exponents is zero. On these rays, one solution transitions from being exponentially large to exponentially small relative to another.

Due to the Stokes phenomenon, no solution of the form \eqref{eq:asymptotics} exists in the neighborhood of an irregular singularity. Consequently, the decomposition in terms of the associator and local monodromy breaks down in this setting. Nevertheless, one can still perform a perturbative analysis using the general expression \eqref{eq:monodromy form}.
\subsection{The scaling transformation} 
The regular KZ connection is invariant under the scaling of the coordinates (conformal invariance), whereas the irregular connection \eqref{eq:irregular connection} is invariant only up to a gauge transformation. Namely, consider the infinitesimal gauge transformation
\begin{equation}
\delta \omega_i = \partial_i \phi - [\phi, \omega_i] ,
\end{equation}
with $\phi=\epsilon\sum_j\omega_jz_j$ where $\epsilon$ is an infinitesimal parameter. This transformation can be interpreted as an infinitesimal scaling of the coordinate $z$:
\begin{equation}
\begin{aligned}
\delta \omega_i
&= \epsilon \Bigl( \omega_i + \sum_j (\partial_i \omega_j)z_j + \sum_j z_j [\omega_j, \omega_i] \Bigr) \\
&= \epsilon \Bigl( 1 + \sum_j z_j \partial_j \Bigr) \omega_i .
\end{aligned}
\end{equation}
For the connection (\ref{eq:irregular connection}) we obtain:
\begin{equation}
\delta \omega_i=\epsilon H_i.
\end{equation}
Therefore, the connection takes the form $\omega_i'=(1+\epsilon)H_i+\ldots$ after the transformation, where $\ldots$ stands for the regular part of the connection, which remains invariant.

We now replace $H_i$ by $\Lambda H_i$ in order to track the scaling behavior of the irregular part of the connection. This computation shows that the resulting connection—and hence its monodromy and the associated braid group representation—is equivalent for any two nonzero values $\Lambda$ and $\Lambda'$.

This means that, in the perturbation theory where all the monodromies are power series in $H_i$ and $\Omega_{ij}$, the link invariant must be independent of $H_i$. Otherwise, rescaling the coordinates --- which does not change the isotopy class of the resulting link --- would lead to different values of the invariant, resulting in a contradiction. An alternative point of view on this is that, when one considers the link obtained by closing a braid as in Figure \ref{fig:braid-closure-solid-torus}, the scale transformation changes the basepoint of the loop in the configuration space of points, but not the homotopy class of the map $S^1\rightarrow Conf_n$.

We can also explicitly check this using the associator $\Psi$. The dependence on the $H_i$'s will drop out if we conjugate the local exponents with $\Psi$ and $\Psi^{-1}$ and take the trace.

However, one can still obtain a new invariant for tangles (rather than the ordinary braids) in the limit where the initial and terminal positions for some of the strands are sent to infinity simultaneously with the limit $\Lambda\rightarrow 0$. We refer to Section \ref{sec:ex} for details.

A single irregular singularity, either at infinity or at a finite point, leads to the same link invariants as in the regular case. However, the scaling transformations of irregular operators associated with singularities at infinity and those located at finite points can cancel each other. Consider an infinitesimal dilatation of the general connection \eqref{eq:irregular connection} (shown above to be equivalent to a gauge transformation):
\begin{equation}
    \begin{aligned}
        \delta\omega_i&=\epsilon(1+\sum_jz_j\partial_j)\omega_i\\
        &=\epsilon\left(1-(l+1)\sum_{j\neq i}\frac{\Omega_{ij}^l}{(z_i-z_j)^{l+1}}\right)+\epsilon(1+l-1)H_i^lz_i^{l-1}\\
        &=\epsilon\sum_l\left(-l\sum_{j\neq i}\frac{\Omega_{ij}^l}{(z_i-z_j)^{l+1}} +lH_i^lz_i^{l-1}\right).
    \end{aligned}
\end{equation}
The scaling behavior depends on the rank $l$. In particular, the operators $\Omega^l$ and the operators at infinity $H^l$ scale with opposite signs. Under the finite transformation $z\rightarrow \lambda z$, we have $H^l\rightarrow H^l/\lambda^l$ and $\Omega_{ij}^l\rightarrow \lambda^l \Omega_{ij}^l$. Consequently, suitable combinations of $\Omega^l$ and $H^l$ can form scale-invariant polynomials, for example $H^1\Omega_{ij}^1$. 

In the following, we will also use capital alphabet $A,B,\ldots$ to denote the general matrices and Lie algebra operators $\Omega_{ij},H$ for brevity. For example, the following equations may give us a non-trivial, irregular-term dependent knot invariant:
\begin{equation}
    \begin{gathered}
        \partial_x\psi=\frac{\Omega^1_{12}}{x^2}+\frac{\Omega_{12}^0}{x}+\frac{\Omega^0_{23}}{x-1}+H,\\
    \partial_x\psi=\frac{A}{x}+\frac{B}{x-1}+\frac{D}{(x-2)^2}+H.
    \end{gathered}
\end{equation}

\section{Examples of Invariants}
\label{sec:ex}
In this section, we do explicit computations of monodromies and the corresponding invariants in various simple settings. Namely,

\begin{itemize}
    \item 1 regular singularity and 1 rank-1 irregular singularity.

\item 2 regular singularities and 1 rank-1 irregular singularity.

\item 2 irregular singularities.
\end{itemize}

\subsection{One regular and one irregular singularity}

\begin{figure}[ht!]
\centering
\begin{tikzpicture}[scale=0.4,baseline=0]
\begin{knot}[ flip crossing=8, flip crossing=2,flip crossing=4, flip crossing=7]

  \strand (-1, 0) node[below]{$0$}  to [out=up, in=down]  (-1,10);
 \strand (-3, 0) node[below]{$x$} 
 to [out=up, in=down] (1,3)
  to [out=up, in=down] (-3,7)
 to [out=up, in=down]  (1,10) node[above]{$-x$};
 \end{knot}
\end{tikzpicture}
\caption{\label{fig: simple braid}}
\end{figure}

Consider two strands with an irregular singularity at infinity. One strand is fixed at the point $0$. The other, with coordinate $z$, can freely move around. The corresponding equation reads
\begin{equation}
\kappa\partial_z\psi=(\frac Az+\Lambda H)\psi.
\end{equation}

We analyze this system using the exact matrix approach. Let $A$ be given by \eqref{eq: matrix rep}, and $H$ by \eqref{eq:irre matrix rep}. We restrict attention to the middle $2\times 2$ block:
\begin{equation}
A=\begin{pmatrix}
-\frac12&1\\
1&-\frac12
\end{pmatrix},\
H=\begin{pmatrix}
-\frac12&0\\
0&\frac12 
\end{pmatrix}.
\end{equation}
A convenient choice of solution basis is given by
\begin{equation}
\begin{aligned}
       \psi_1(z)&=\begin{pmatrix}
        z^{\frac1{2\kappa}}\exp(-\frac{z\Lambda}{2\kappa})U(\frac1\kappa,1+\frac2\kappa,\frac{z\Lambda}{\kappa})\\
        \frac12(1+z\Lambda+2z\kappa\partial_z)z^{\frac1{2\kappa}}\exp(-\frac{z\Lambda}{2\kappa})U(\frac1\kappa,1+\frac2\kappa,\frac{z\Lambda}{\kappa})
    \end{pmatrix},\\
      \psi_2(z)&=\begin{pmatrix}
        z^{\frac1{2\kappa}}\exp(-\frac{z\Lambda}{2\kappa}){}_1F_1(\frac1\kappa,1+\frac2\kappa,\frac{z\Lambda}{\kappa})\\
        \frac12(1+z\Lambda+2z\kappa\partial_z)z^{\frac1{2\kappa}}\exp(-\frac{z\Lambda}{2\kappa}){}_1F_1(\frac1\kappa,1+\frac2\kappa,\frac{z\Lambda}{\kappa})
    \end{pmatrix} .
\end{aligned}
\end{equation}
Here and below $U$ denotes Tricomi confluent hypergeometric function and $_pF_q$ denotes generalized hypergeometric function. Apart from the prefactor $z^{\frac1{2\kappa}}$, the functions $\psi_1$ and $\psi_2$ depend on $z$ and $\Lambda$ solely through the product $z\Lambda$.
We may rewrite the solution basis in the form
\begin{equation}
(\psi_1(z),\psi_2(z))=z^{\frac{1}{2\kappa}}\exp(-\frac{z\Lambda}{2\kappa})\tilde{\psi}(z\Lambda).
\end{equation}
Here $\tilde{\psi}(z\Lambda)$ denotes a matrix-valued function depending only on the combination $z\Lambda$.

From the theory of special functions \cite{abramowitz+stegun}, the monodromy matrix $M$ is given by 
 \begin{equation}
     \begin{pmatrix}
         \psi_1(e^{2\pi i}z)&\psi_2(e^{2\pi i}z)
     \end{pmatrix}=\begin{pmatrix}
         \psi_1(z)&\psi_2(z)
     \end{pmatrix}\begin{pmatrix}
         \exp(\frac{-3\pi i}{\kappa})&0\\
         (\exp(\frac{\pi i}{\kappa})-\exp(\frac{-3\pi i}{\kappa}))\frac{\Gamma(-\frac2\kappa)}{\Gamma(-\frac1\kappa)}&\exp(\frac{\pi i}{\kappa})
     \end{pmatrix}.
 \end{equation}

As discussed in Section 2, we also consider configurations that start at the point $x$, wind once around the strand at $0$, and end at $-x$ (see Figure~\ref{fig: simple braid}). The corresponding parallel transport from $z=x$ to $z=-x$ is not a topological quantity, as it depends explicitly on the value of $x$. Nevertheless, it still defines a linear transformation on the vector (fiber) space, which can be computed as 
\begin{equation}
\begin{aligned}
   P'=&\begin{pmatrix}
        \psi_1(-x)&\psi_2(-x)
    \end{pmatrix}\begin{pmatrix}
        \psi_1(x)&\psi_2(x)
    \end{pmatrix}^{-1}\\
    =&(-1)^{\frac1{2\kappa}}\exp(\frac{x\Lambda}{\kappa})\tilde{\psi}(-x\Lambda,\kappa)\tilde{\psi}^{-1}(x\Lambda,\kappa).
    \end{aligned}
\end{equation}
Taking the limits $\Lambda\rightarrow 0$ and $x\rightarrow \infty$, while keeping the product $x\Lambda=\lambda$ fixed, we obtain a constant (i.e., independent of the positions of the strand endpoints) matrix.

To obtain a basis-independent tangle invariant, we can consider the trace of the matrix product. For example 
\begin{equation}
\begin{gathered}
      \mathrm{Tr}(P')=\frac{\text{Numerator}}{\text{Denominator}},\\
      \text{Numerator}=q^{-\frac14}e^{\frac t2}\sqrt{-t}(1+\frac2\kappa)K(\frac12+\frac1\kappa,-t){}_1F_1(1+\frac1\kappa,2+\frac2\kappa,t)\\+q^{\frac14}e^{\frac{3t}{2}}\sqrt t(1+\frac2\kappa)K(\frac12+\frac1\kappa,t){}_1F_1(1+\frac1\kappa,2+\frac2\kappa,-t)\\+\sqrt\pi q^{\frac14}e^tt^{1+\frac1\kappa}[{}_1F_1(\frac1\kappa,1+\frac2\kappa,t)U(\frac1\kappa,1+\frac2\kappa,-t)-{}_1F_1(\frac1\kappa,1+\frac2\kappa,-t)U(\frac1\kappa,1+\frac2\kappa,t)],\\
      \text{Denominator}=
      e^{\frac t2}\sqrt{t}(\frac2\kappa+1)K(\frac12+\frac1\kappa,\frac t2){}_1F_1(\frac1\kappa,1+\frac2\kappa,t)\\+\sqrt{\pi} t^{1+\frac1\kappa}{}_1F_1(1+\frac1\kappa,2+\frac2\kappa,t)U(\frac1\kappa,1+\frac2\kappa,t),
\end{gathered}
\end{equation}
 where $K$ denotes the modified Bessel function of the second kind. For convenience, we introduce the notation $t=\frac{\lambda}{\kappa}$ and $q=\exp(\frac{2\pi i}{\kappa})$.

\subsection{Two regular singularities and one irregular}
Consider three strands with an irregular singularity at infinity. We fix the positions of two strands at $0$ and $1$ corresponding to two regular singularities, and allow the third strand, with coordinate $z$, to go freely around them. More specifically, we consider the following matrix differential equation:
\begin{equation}
    \kappa\partial_z\psi=\left[\frac{1}{z}\begin{pmatrix}
        0&0&0\\
        0&1&-1\\
        0&-1&1
    \end{pmatrix}+\frac{1}{z-1}\begin{pmatrix}
        1&-1&0\\
        -1&1&0\\
        0&0&0
    \end{pmatrix}+
    \Lambda\begin{pmatrix}
        0&0&0\\
        0&-1&0\\
        0&0&0
    \end{pmatrix}\right]\psi.
\end{equation}
This equation admits the following basis of integral solutions \cite{gu2023irregularfibonacciconformalblocks}:
\begin{equation}
    \begin{aligned}
        \psi_1=(\frac\Lambda\kappa)^{\frac3\kappa}\int_{0}^z\begin{pmatrix}
            w^\frac1\kappa(w-z)^\frac{1}{\kappa}(w-1)^{\frac{1}{\kappa}-1}\exp(-\frac{\Lambda}{\kappa}w),\\
            w^\frac{1}{\kappa}(w-z)^{\frac{1}{\kappa}-1}(w-1)^{\frac{1}{\kappa}}\exp(-\frac{\Lambda}{\kappa}w),\\
            w^{\frac{1}{\kappa}-1}(w-z)^{\frac{1}{\kappa}}(w-1)^{\frac{1}{\kappa}}\exp(-\frac{\Lambda}{\kappa}w)
        \end{pmatrix}dw,\\
    \end{aligned}
\end{equation}

\begin{equation}
   \begin{aligned}
        \psi_2=(\frac\Lambda\kappa)^{\frac3\kappa}\int_{z}^1\begin{pmatrix}
            w^\frac1\kappa(w-z)^\frac{1}{\kappa}(w-1)^{\frac{1}{\kappa}-1}\exp(-\frac{\Lambda}{\kappa}w)\\
            w^\frac{1}{\kappa}(w-z)^{\frac{1}{\kappa}-1}(w-1)^{\frac{1}{\kappa}}\exp(-\frac{\Lambda}{\kappa}w)\\
            w^{\frac{1}{\kappa}-1}(w-z)^{\frac{1}{\kappa}}(w-1)^{\frac{1}{\kappa}}\exp(-\frac{\Lambda}{\kappa}w)
        \end{pmatrix}dw
    \end{aligned},
\end{equation}

\begin{equation}
    \begin{aligned}
\psi_3=\frac{(\frac\Lambda\kappa)^{\frac3\kappa}}{\Gamma(\frac3\kappa)}\int_{1}^{\infty}\begin{pmatrix}
            w^\frac1\kappa(w-z)^\frac{1}{\kappa}(w-1)^{\frac{1}{\kappa}-1}\exp(-\frac{\Lambda}{\kappa}w)\\
            w^\frac{1}{\kappa}(w-z)^{\frac{1}{\kappa}-1}(w-1)^{\frac{1}{\kappa}}\exp(-\frac{\Lambda}{\kappa}w)\\
            w^{\frac{1}{\kappa}-1}(w-z)^{\frac{1}{\kappa}}(w-1)^{\frac{1}{\kappa}}\exp(-\frac{\Lambda}{\kappa}w)
        \end{pmatrix}dw.
    \end{aligned}
\end{equation}
In principle, one could repeat the computation of the monodromy and the tangle matrix as in the previous section. However, the resulting explicit expressions are rather intricate, and we will not pursue this approach here. Instead, we show that the solution basis $\psi_i(\Lambda,x)$ remains finite in the limit $x\rightarrow \infty, \Lambda\rightarrow0$, with $ x\Lambda=\lambda$ held fixed.

As an example, consider a component of $\psi_1$:
\begin{equation}
    \Lambda^{\frac3\kappa}\int_{0}^z
            w^\frac1\kappa(w-z)^\frac{1}{\kappa}(w-1)^{\frac{1}{\kappa}-1}\exp(-\frac{\Lambda}{\kappa}w)dw.
\end{equation}
Let $y=w/z$. Changing the integration variable and substituting $\Lambda\rightarrow \frac{\lambda}{z}$, we obtain:
\begin{equation}
\begin{aligned}
     &\Lambda^{\frac3\kappa}\int_{0}^1
            (zy)^\frac1\kappa(zy-z)^\frac{1}{\kappa}(zy-1)^{\frac{1}{\kappa}-1}\exp(-\frac{\Lambda}{\kappa}zy)zdy\\
            =&(\frac{\lambda}{z})^{\frac3\kappa}z^{\frac3\kappa}\int_0^1 y^{\frac1\kappa}(y-1)^{\frac1\kappa}(y-\frac1z)^{\frac1\kappa-1}\exp(-\frac\lambda\kappa y)dy.\\
\end{aligned}
\end{equation}
In the limit $z\rightarrow \infty$, we have $\frac 1 z\rightarrow 0$, and the above expression reduces to
\begin{equation}
    (-1)^\frac1\kappa\lambda^\frac3\kappa\Gamma(\frac1\kappa+1)\Gamma(\frac2\kappa){}_1F_1(\frac2\kappa,\frac3\kappa+1,-\frac\lambda\kappa).
\end{equation}
Similarly we can verify that the other elements converge under this limit.

Now we can see that the holonomy matrix from $z=x$ to $z=-x$ 
\begin{equation}
    (
    (\psi_1(-x),\psi_2(-x),\psi_3(-x))(\psi_1(x),\psi_2(x),\psi_3(x))^{-1}
\end{equation}
is well-defined under the above limits. That implies the tangle invariant is well-defined. 

\subsection{Two irregular singularities and one irregular: universal perturbative invariant}
\begin{figure}[ht!]
\centering
\begin{tikzpicture}[scale=0.4,baseline=0]
\begin{knot}[flip crossing=1, flip crossing=8, flip crossing=2,flip crossing=4, flip crossing=10, flip crossing=7, flip crossing=12, flip crossing=14]

  \strand (-1, 0) node[below]{$0$}  to [out=up, in=down]  (-1,25);
 \strand (1, 0) node[below]{$1$}  to [out=up, in=down]  (1,25);
 \strand (-5, 0) node[below]{$x$} 
 to [out=up, in=down] (2,6)
  to [out=up, in=down] (-2,8)
   to [out=up, in=down] (0,10)
   to [out=up, in=down] (-2,11)
   to [out=up, in=down] (2,13)
   to [out=up, in=down] (-2,15)
    to [out=up, in=down] (2,17)
    to [out=up, in=down] (0,19)
 to [out=up, in=down]  (6,25) node[above]{$-x$};
\end{knot}
\end{tikzpicture}
\caption{\label{fig:3-strands-special}}
\end{figure}
Consider the special case when the two strands are fixed at 0 and 1, while the third starts at $x$ and ends at $-x$, as in Figure \ref{fig:3-strands-special}. That is we are interested in the holonomy
\begin{equation}
    P\,\exp\int\limits_{x}^{-x}  dz \left\{\Lambda\,H+\frac{A}{z} 
    +\frac{B}{z-1}\right\}
\end{equation}
where the integral is along the path corresponding to the third strand. In the limit $\Lambda\rightarrow 0$, $x\rightarrow \infty$, $\Lambda x\rightarrow 1$ its perturbative expansion in $A,B,H$ reads:
\begin{multline}
    1+2H+\left(\pi i\,n_0\right)A+\left(\pi i\,n_1\right)B+\\
    +\left(\pi i\,n_0-2\right)HA
    +\left(\pi i\,n_0+2\right)AH
    +\left(\pi i\,n_1-2\right)HB
    +\left(\pi i\,n_1+2\right)BH+\\
    -\frac{\pi^2n_0^2}{2}A^2-\frac{\pi^2n_1^2}{2}B^2+\ldots\\
    +\left(-2-\frac{\pi i}{2}n_0\right)AHAH+\left(-2+\frac{\pi i}{2}n_0\right)HAHA+\ldots
\end{multline}
where $n_0$ and $n_1$ are the total numbers of half-turns of the third strand around 0 and 1 respectively. In general, the coefficients are invariants that cannot be expressed through $n_0,n_1$ only. Note that taking the formal trace operation, as in (\ref{2irr-trace-expansion}) cannot be used to remove $H$ dependence:
\begin{multline}
    \mathrm{Tr}\,1+2\,\mathrm{Tr}\,H+\left(\pi i\,n_0\right)\mathrm{Tr}\,A+\left(\pi i\,n_1\right)\mathrm{Tr}\,B
    +\left(2\pi i\,n_0\right)\mathrm{Tr}\,HA
        +\left(2\pi i\,n_1\right)HB
   \\
    -\frac{\pi^2n_0^2}{2}\mathrm{Tr}\,A^2-\frac{\pi^2n_1^2}{2}\mathrm{Tr}\,B^2+\ldots
    +\left(-2{\pi i}n_0\right)\mathrm{Tr}\,HAHA+\ldots 
\end{multline}
Taking the trace again corresponds to considering the tangle in $D^2\times S^1$ as in Figure \ref{fig:tangle-closure-solid-torus}. Alternatively, one can consider the tangle in $\mathbf{C}\times S^1$, with one of the strands understood as a non-compact one, going to $+\infty$ and $-\infty$ in $\mathbf{C}$ at fixed positions on $S^1$. 
\subsection{Two irregular operators}

We consider a configuration with two irregular singularities of degree $1$ at zero and infinity, together with one degenerate (spin $\frac{1}{2}$) operator at $z$. Here, we will analyze the monodromy of transporting the regular operator around the irregular singularity at $z=0$ using the Stokes phenomenon. The conformal block or wavefunction we are interested in is of the form
\begin{equation}
    \psi(z) = \langle I_{\infty}^{(1)}| \Phi(z) | I_0^{(1)}\rangle,
\end{equation}
where $\Phi(z)$ is the degenerate spin $\frac{1}{2}$ field and $I^{(1)}$ are irregular operators of degree $1$. The KZ connection takes the form
\begin{equation}
    \nabla_z = \kappa \partial_z - A(z),
\end{equation}
where $A(z)$ is of the form 
\begin{align}
    A(z) &= \Omega_{\infty}^{(1)} + \frac{1}{z} \Omega^{(0)}_0 + \frac{1}{z^2} \Omega^{(1)}_0 \nonumber \\
    ~ &= \left(
    \begin{array}{cccc}
        v + \frac{1}{2z} + \frac{u}{z^2} & 0 & 0 & 0 \\
        0 & v - \frac{1}{2z} + \frac{u}{z^2} & \frac{1}{z} & 0 \\
        0 & \frac{1}{z} & - v - \frac{1}{2z} - \frac{u}{z^2} & 0 \\
        0 & 0 & 0 & - v + \frac{1}{2z} - \frac{u}{z^2}
    \end{array}     
    \right)~.
\end{align}
We thus see that the non-abelian dynamics and non-trivial Stokes phenomena entirely reside in the $2 \times 2$ block
\begin{equation}
    A_{\textrm{core}}(z) = \left(
    \begin{array}{cc}
        v - \frac{1}{2z} + \frac{u}{z^2} & \frac{1}{z} \\
        \frac{1}{z} & -v - \frac{1}{2z} - \frac{u}{z^2}
    \end{array}
    \right)~.
\end{equation}
In order to study monodromy properties and Stokes phenomena of this connection, we decompose
\begin{align}
    A_{\textrm{core}}(z) &= W + \frac{R}{z} + \frac{D}{z^2} \nonumber \\
    ~ &= \left(\begin{array}{cc}
        v & 0 \\
        0 & -v
    \end{array}\right) + \frac{1}{z} \left(\begin{array}{cc}
        -1/2 & 1 \\
         1   & -1/2
    \end{array}\right) + \frac{1}{z^2} \left(\begin{array}{cc}
        u & 0 \\
        0 & -u
    \end{array}\right)~.
\end{align}
Then the formal solution of $\nabla_{z,\textrm{core}} \psi = 0$ takes the form
\begin{equation}
\label{eq:formal basis}
    \psi_{\textrm{formal}}(z) = \hat F(z) z^{\Lambda/\kappa}e^{Q(z)/\kappa},
\end{equation}
where the irregular exponent $Q(z)$ and the formal monodromy $\Lambda$ are dictated by the diagonal entries:
\begin{equation}
Q(z) = \int \frac{D}{z^2}dz = \left(\begin{array}{cc}
        -u/z & 0 \\
         0   & u/z
    \end{array}\right), \quad
    \Lambda = \textrm{diag}(R) = \left(\begin{array}{cc}-1/2 & 0\\ 0 & -1/2\end{array}\right)~.
\end{equation}
Plugging this ansatz into the equation $\nabla_{z,\textrm{core}} \psi_{\textrm{formal}}= 0 $ then gives a recursive solution to the coefficients in 
\begin{equation}
    \hat F(z) = \sum_{k=0}^{\infty} F_k z^k.
\end{equation}
The Stokes multipliers are then determined via
\begin{align}
    s_1&= \lim_{k\rightarrow \infty} \frac{2\pi i \cdot (2u/k)^k}{\Gamma(k)}(F_k)_{12} \label{eq:s1lim}, \\
    s_2 &= \lim_{k\rightarrow \infty} \frac{2\pi i \cdot (-2 u/\kappa)^k}{\Gamma(k)}(F_k)_{21} \label{eq:s2lim}.
\end{align}
The total monodromy around $z=0$ is then given by
\begin{equation}
    M_0 = S_1 S_2 e^{2\pi i \Lambda/\kappa}, \quad S_1 = \left(\begin{array}{cc}1 & s_1\\ 0 & 1\end{array}\right), \quad S_2 = \left(\begin{array}{cc}1 & 0\\ s_2 & 1\end{array}\right).
\end{equation}
By matrix multiplication
\begin{equation}
    M_0=\begin{pmatrix}
        1+s_1s_2&s_1\\
        s_2&1
    \end{pmatrix}\begin{pmatrix}
        e^{-\pi i/\kappa}&0\\
        0&e^{-\pi i/\kappa}
    \end{pmatrix}
\end{equation}

Instead of computing the limits \eqref{eq:s1lim} and \eqref{eq:s2lim} numerically, we will solve for them analytically in the semiclassical limit $\kappa \rightarrow 0$. To this end, note that $A_{\textrm{core}}(z)$ is not traceless but can be made traceless via the gauge transformation $\psi = z^{-1/(2\kappa)} \tilde \psi$. This removes the $- \frac{1}{2z}$ term, leaving the trace-free matrix:
\begin{equation}
    \tilde A(z) = \left(\begin{array}{cc}
       v + \frac{u}{z^2}  & \frac{1}{z} \\
        \frac{1}{z} & -v - \frac{u}{z^2}
    \end{array}\right).
\end{equation}
We can then define the spectral curve $\Sigma$ associated to $\tilde A$ via the characteristic equation $\textrm{det}(\tilde A(z) - \lambda I) = 0$:
\begin{equation}
    \lambda^2 - \left[\left(v + \frac{u}{z^2}\right)^2 + \frac{1}{z^2}\right] = 0.
\end{equation}
This defines a double cover of the punctured sphere, $\Sigma : \lambda^2 = \phi_2(z)$, where the quadratic differential $\phi_2(z) dz^2$ is
\begin{equation}
    \phi_2(z) dz^2 = \left(v^2 + \frac{2uv + 1}{z^2} + \frac{u^2}{z^4}\right)dz^2.
\end{equation}
A WKB analysis then shows that to leading order, the analytic values of the Stokes multipliers are given by the exponentiated periods along specific cycles $\gamma_i$ of the curve $\Sigma$:
\begin{equation}
    s_i \sim \exp\left(\frac{1}{\kappa} \oint_{\gamma_i} \sqrt{\phi_2(z)}\right).
\end{equation}
The critical points of the WKB analysis are the turning points where $\phi_2(z) = 0$. These are the branch points of the spectral curve, obtained by setting the numerator of $\phi_2(z)$ to zero:
\begin{equation}
    v^2 z^4 + (2uv + 1)z^2 + u^2 = 0.
\end{equation}
This is a quadratic equation in $z^2$, the four turning points are $z= \pm \sqrt{x_+}$, $\pm \sqrt{x_-}$, where
\begin{equation}
    x_\pm = \frac{-(2 u v +1) \pm \sqrt{4uv + 1}}{2v^2}.
\end{equation}
The branch cuts connecting these turning points define the basic cycles $\gamma_i$ on the Seiberg-Witten curve. The Stokes lines of the exact WKB method are the trajectories emanating from these turning points, defined by the condition:
\begin{equation}
    \textrm{Im}\int^z \sqrt{\phi_2(w)}dw = 0.
\end{equation}
Let the period integrals along the respective BPS cycles be:
\begin{equation}
    \Pi_1 = \oint_{\gamma_1} \sqrt{\phi_2(w)}dw, \quad \Pi_2 = \oint_{\gamma_2} \sqrt{\phi_2(w)}dw. 
\end{equation}
The semiclassical Stokes multipliers are $s_1 \sim e^{\Pi_1/\kappa}$ and $s_2 \sim e^{\Pi_2/\kappa}$. 
Geometrically, the concatenation of the two Stokes cycles $\gamma_1 + \gamma_2$ forms a single, closed homology cycle $\gamma_0$ on the Seiberg-Witten curve $\Sigma$ that completely encloses the irregular puncture at $z=0$. This means that the dynamical term $s_1 s_2$ is governed by the period of the quadratic differential around the origin:
\begin{equation}
    s_1 s_2 \sim \exp\left(\frac{1}{\kappa} \oint_{\gamma_0}\sqrt{v^2 + \frac{2uv + 1}{w^2} + \frac{u^2}{w^4}}dw\right).
\end{equation}
To evaluate the loop integral $\oint_{\gamma_0} \sqrt{\phi_2(w)}dw$, we can compute it using Cauchy's residue theorem by studying the behavior of the quadratic differential near the singularity at $w=0$. To find the residue at $w=0$, we factor out the dominant singular term to expand the square root in a Laurent series:
\begin{equation}
    \sqrt{\phi_2(w)} = \sqrt{\frac{u^2}{w^4}\left(1 + \frac{2uv+1}{u^2}w^2 + \frac{v^2}{u^2 }w^4\right)} = \frac{u}{w^2}\sqrt{1 + Aw^2 + B w^4},
\end{equation}
where we have defined constants $A = \frac{2uv+1}{u^2}$ and $B = \frac{v^2}{u^2}$. Taylor expanding the square root term, we find that
\begin{equation}
    \textrm{Res}_{w=0} \sqrt{\phi_2(w)} = 0.
\end{equation}
Thus we find
\begin{equation}
    \Pi_1 + \Pi_2 = 0 \quad \Rightarrow \quad \Pi_2 = - \Pi_1.
\end{equation}
Because the semiclassical Stokes multipliers are given by $s_i \sim \exp\left(\Pi_2/\kappa\right)$, this forces the two multipliers to be exact inverses of each other at leading order:
\begin{equation}
    s_2 \sim s_1^{-1}~.
\end{equation}

Inserting these into the exact expression for $M_0$ gives the semiclassical matrix:
\begin{equation}
    M_0^{(sc)} \sim \left(\begin{array}{cc}
        1 + \exp(\frac{\Pi_1 + \Pi_2}{\kappa}) & \exp(\frac{\Pi_1}{\kappa})  \\
        \exp(\frac{\Pi_2}{\kappa}) & 1 
    \end{array}
    \right)\begin{pmatrix}
        e^{-\frac{\pi i}{\kappa}}&0\\
        0&e^{-\frac{\pi i}{\kappa}}
    \end{pmatrix}.
\end{equation}
As a consequence, the trace of the semiclassical monodromy $M_0^{(sc)}$ evaluates to 
\begin{equation}
    \textrm{Tr}\left[M_0^{(sc)}\right] \sim e^{-\frac{\pi i}{\kappa}}\left(1 + \exp\left(\frac{\Pi_1 + \Pi_2}{\kappa}\right)\right)+e^{-\frac{\pi i}{\kappa}}=3e^{-\frac{\pi i}{\kappa}}.
\end{equation}
Note that, our WKB basis may differ from the one in \eqref{eq:formal basis} by a linear transformation. This may affect the monodromy matrix but not the trace. The only other basis-independent invariant is $\det{M_0^{(sc)}}=e^{-\frac{2\pi i}{\kappa}}$.

Next, we compute the period $\Pi_1$ perturbatively. The natural dimensionless expansion parameter for this geometry is the coupling $g = uv$, hence we aim to expand the integral in powers of small $g$. Changing variables to $x = z^2$ (and hence $dz = \frac{dx}{2\sqrt{x}}$), the integral becomes:
\begin{equation}
    \Pi_1 = \oint_{\tilde \gamma_1} \frac{\sqrt{v^2 x^2 + (2uv+1)x + u^2}}{2x^{3/2}}dx.
\end{equation}
The roots of the polynomial in the numerator are given by
\begin{equation}
    x_\pm = \frac{-(2g+1)\pm \sqrt{1+4g}}{2v^2}.
\end{equation}
For small $g$, $x_+$ is close to the origin ($x_+ \approx -u^2$), and $x_-$ is near infinity ($x_- \approx -1/v^2$). The cycle $\tilde \gamma_1$ corresponds to the branch cut between $0$ and $x_+$. By factoring the quadratic differential as $v^2(x - x_+)(x-x_-)$, we can evaluate the contour integral in terms of the Gauss Hypergeometric function:
\begin{equation}
    \Pi_1 = C \cdot v \sqrt{-x_-} \cdot {}_{2}F_{1}\left(-\frac{1}{2}, - \frac{1}{2};1;\frac{x_+}{x_-}\right),
\end{equation}
where $C$ is a constant phase. Finally, using the Taylor expansions
\begin{align}
    {}_{2}F_{1}(-1/2,-1/2;1;w) &= 1 + \frac{1}{4}w + \frac{1}{64}w^2 + \ldots, \\
    \frac{x_+}{x_-} &= g^2 - 4 g^3 + 14 g^4 + \mathcal{O}(g^5)
\end{align}
gives
\begin{equation}
    s_1 \sim \exp\left(\frac{\Pi_1}{\kappa}\right) = \exp\left[\pm \frac{i\pi}{\kappa}\left(1 + uv - \frac{3}{4} u^2 v^2 + \frac{5}{4}u^3 v^3 + \ldots \right)\right].
\end{equation}
\subsection{Two irregular singularities: universal perturbative invariant}

\begin{figure}[ht!]
\centering
\begin{tikzpicture}[scale=0.4,baseline=0]
\begin{knot}[flip crossing=1, flip crossing=8, flip crossing=2,flip crossing=4, flip crossing=7]

  \strand (-1, 0) node[below]{$0$}  to [out=up, in=down]  (-1,25);
 \strand (-5, 0) node[below]{$x$} 
 to [out=up, in=down] (2,6)
  to [out=up, in=down] (-2,8)
   to [out=up, in=down] (0,10)
   to [out=up, in=down] (-2,11)
   to [out=up, in=down] (2,13)
   to [out=up, in=down] (-2,15)
    to [out=up, in=down] (2,17)
 to [out=up, in=down]  (-5,25) node[above]{$y$};
\end{knot}
\end{tikzpicture}
\caption{\label{fig:2-irr-sing-1-reg-braid}}
\end{figure}

The case of the two irregular singularities can also be considered on the formal level. Namely, one can calculate
\begin{equation}
    P\exp\int_x^y  dz \left\{H+\frac{A}{z} 
    +\frac{F}{z^2}\right\}
\end{equation}
as a formal series in the non-commutative variables $A,H,F$. The integral above is performed under some path avoiding zero. The path corresponds to a strand in $\mathbb{C}\times \mathbb{R}$ going around the straight strand supported at $\{0\}\times \mathbb{R}$ corresponding to the irregular singularity at $0$, as shown in Figure \ref{fig:2-irr-sing-1-reg-braid}.   In this case, there are no relations coming from the flatness condition. Moreover, in order to get a purely topological invariant, we will set $y=x$ and consider a formal trace operation, which makes equivalent the words in $A,H,F$ that can be related by a cyclic permutation:
\begin{multline}
   \mathrm{Tr}\, P\exp\int\limits_{x}^x  dz \left\{H+\frac{A}{z} 
    +\frac{F}{z^2}\right\}=\\
    =\mathrm{Tr}\,e^{2\pi inA}+2(\pi in)^2\,\left(\mathrm{Tr}\,AFH-\mathrm{Tr}\,AHF\right)+4(\pi in)^2\,\mathrm{Tr}\,FHFH+8(\pi in)^2\,\mathrm{Tr}\,AHAF\\
    -4\left[(\pi in)^2+(\pi in)^3\right]\,\mathrm{Tr}\,AAHF-4\left[(\pi in)^2-(\pi in)^3\right]\,\mathrm{Tr}\,AAFH+\ldots
    \label{2irr-trace-expansion}
\end{multline}
Where $n\in \mathbb{Z}$ is the signed number of times the non-fixed strand winds around the strand fixed at zero. The trace operation corresponds to closing the braid in $\mathbb{C}\times S^1$.

\appendix
\section*{Acknowledgements}
We would like to thank N. Reshetikhin for valuable discussions. XG would like to thank Yong Li and Xin-Xing Tang for discussions during the early stages of this work. PP would like to thank BIMSA and YMSC, where the collaboration on the project started, for their hospitality. The work of BH was supported by NSFC grant 2250610187.
\section{The associator $\Psi$}
\label{A:associator}
In this appendix, we describe the procedure for expressing the associator $\Psi$ in terms of the generators $\Omega_{ij}$ and compute its matrix representation for exact solutions. 
We begin with the differential equation:
\begin{equation}
\label{eq:reduced KZ}
    \kappa\partial_x\psi=(\frac{A}{x}+\frac{B}{x-1})\psi.
\end{equation}
This equation can be viewed as a special case of the $KZ$ equation of $n=3$, obtained by fixing $z_1=0$ and $z_3=1$ while retaining the dependence on $z_2=x$. The full coordinate dependence can be recovered by applying conformal transformations.

Here $A$ and $B$ play the roles of $\Omega_{12}$ and $\Omega_{23}$, respectively. For the purposes of this discussion, we regard them simply as two noncommuting abstract variables, and we seek solutions in the ring of formal power series $\mathbf{C}[[A,B]]$. We have also introduced a parameter $\kappa$ to keep track of the degree in the $\Omega_{ij}$.  

Equation \eqref{eq:reduced KZ} is an ODE with $3$ regular singular points located at $x=0,1$ and $\infty$. The fundamental group $\pi_1$ is generated by loops encircling $0$ and $1$. Consequently, the monodromy matrix $M_\gamma$ associated with any loop $\gamma$ is completely determined by the monodromies around $0$ and $1$.

To define the associator properly, we must specify bases of solutions near $x=0$ and $x=1$. Since the singularities are regular, the situation is straightfoward: there exist unique solutions $\psi_0(x)$ and $\psi_1(x)$ with the following asymptotic behaviour:
\begin{equation}
\label{eq:reduasymptotics}
\begin{aligned}
     \psi_0(x)&\sim x^A \quad x\rightarrow 0,\\
     \psi_1(x)&\sim (1-x)^B \quad x\rightarrow 1.
\end{aligned}
\end{equation}
Here the notation $\sim$ means that 
\begin{equation}
\begin{aligned}
     \psi_0(x)&=f(x)x^A,\\
    \psi_1(x)&=g(1-x)(1-x)^B ,
\end{aligned}
\end{equation}
where $f(x)$ and $g(x)$ are analytic functions in a neighbourhood of $0$ with values in $\mathbf{C}[[A,B]]$ satisfying $f(0)=g(0)=1$.

From these asymptotics, we immediately read off the monodromy:
\begin{equation}
\begin{aligned}
      \psi_0(e^{2\pi i}x)&=\psi_0(x)e^{2\pi iA},\\
      \psi_1(e^{2\pi i}(1-x))&=\psi_1(1-x)e^{2\pi iB}.\\
\end{aligned}
\end{equation}

We define the associator by $\Psi=\psi_1^{-1}\psi_0$. Formally, this involves taking the inverse of one solution and multiplying it by the other. However, the inverse is to be understood in the sense of an order-by-order cancellation in the formal power series expansion.

Any element in $\mathbf{C}[[A,B]]$ can be written as a sum of words of $A$ and $B$, where a words mean monomials such as $ABAB,B^2A$ or $BA$. Since $A$ and $B$ do not commute (for example, $BA\neq AB$), the ordering of letters matters.

We may formally expand $\psi$ as series of $A$ and $B$ because it takes values in $\mathbf{C}[[A,B]]$:
\begin{equation}
\label{eq:ansatz}
\begin{gathered}
        \psi(x)=1+\frac1\kappa (f_A(x)A+f_B(x)B)+\\
        \frac{1}{\kappa^2}(f_{A^2}(x)A^2+f_{AB}(x)AB+f_{BA}(x)BA+f_{B^2}(x)B^2)+\ldots
\end{gathered}
\end{equation}
where all coefficients $f_\bullet(x)$ are scalar functions of $x$. This expression is simultaneously a formal expansion in the generators $A,B$ and an expansion in powers of $\frac1\kappa$.  The length $m$ of a word in $A$ and $B$ always matches the order $\frac1{\kappa^m}$. 

We now substitute the \eqref{eq:ansatz} into equation\eqref{eq:reduced KZ}. By comparing both the powers of $\frac1\kappa$ and the coefficients of words in $A$ and $B$ on the two sides of the equation, we obtain a hierarchy of differential equations:
\begin{equation}
\begin{gathered}
      \partial_xf_A=\frac1x,\partial_xf_B=\frac{1}{x-1},\\
      \partial_xf_{A^2}=\frac {f_A}x,\partial_xf_{AB}=\frac{f_B}{x},\partial_xf_{BA}=\frac{f_A}{x-1},\partial_xf_{B^2}=\frac{f_B}{x-1},\\
      \ldots
\end{gathered}
\end{equation}
We omit the higher-order equations corresponding to words of length $m>2$. These functions $f_\bullet(x)$ can be solved recursively, yielding
\begin{equation}
    \begin{gathered}
        f_A=\log x+C_A,\quad f_B=\log(1-x)+C_B,\\
        f_{A^2}=\frac12(\log x)^2+C_A\log x+C_{A^2},\\
        f_{AB}=\mathrm{Li}_2(1-x)+\log x\log(1-x)+C_B\log x+C_{AB},\\
    f_{BA}=-\mathrm{Li}_2(1-x)+C_A\log(1-x)+C_{BA},\\
    f_{B^2}=\frac12(\log (1-x))^2+C_B\log(1-x)+C_{B^2},\\
    \ldots
    \end{gathered}
\end{equation}
where the $C_
\bullet$ are integration constants. $\mathrm{Li}_2$ denotes the polylogarithm function.

These constants are fixed by imposing the asymptotic boundary condition \eqref{eq:asymptotics}. Near $x=0$, the solution satisfies $\psi_0\sim x^A$, which implies $\lim_{x\rightarrow 0}\psi_0=1+\log x A+\frac12(\log x)^2A^2+\ldots$ To reproduce this behavior, we set $C_A=C_B=C_{A^2}=C_{B^2}=0,$ $C_{AB}=-\frac{\pi^2}{6}$,$C_{BA}=\frac{\pi^2}{6}$.
With these choices, the solution $\psi_0$ takes the form
\begin{equation}
\begin{gathered}
      \psi_0=1+\frac1\kappa(\log x A+\log(1-x)B)+\frac{1}{\kappa^2}(\frac12(\log x)^2A^2+\\
      (\mathrm{Li}_2(1-x)+\log x\log(1-x)-\frac{\pi^2}{6})AB+\\
      (-\mathrm{Li}_2(1-x)+\frac{\pi^2}{6})BA+\frac12(\log(1-x))^2 B^2)+\ldots
\end{gathered}
\end{equation}

For the solution $\psi_1$, the integration constants are fixed as $C_A=C_B=C_{A^2}=C_{AB}=C_{BA}=C_{BB}=0.$ The corresponding solution is therefore
\begin{equation}
\begin{gathered}
       \psi_1=1+\frac1\kappa(\log x A+\log(1-x)B)+\frac1{\kappa^2}(\frac12(\log x)^2A^2+\\
       (\mathrm{Li}_2(1-x)+\log x\log(1-x))AB+\\-\mathrm{Li}_2(1-x)BA
       +\frac12(\log (1-x))^2B^2)+\ldots
\end{gathered}
\end{equation}

The inverse of $\psi_1$ can also be computed order by order, yielding:
\begin{equation}
\begin{gathered}
     \psi_1^{-1}=1-\frac1\kappa(\log x A+\log(1-x)B)+\frac1{\kappa^2}(\frac12(\log x)^2A^2
       -\mathrm{Li}_2(1-x)AB\\+(\mathrm{Li}_2(1-x)+\log x \log(1-x))BA
       +\frac12(\log (1-x))^2B^2)+\ldots   
\end{gathered}
\end{equation}
More generally, the coefficient of order $k$ in $\psi_1^{-1}$ is determined recursively by
\begin{equation}
    \psi_1^{-1}\big|_{k}=-\sum_{j=1}^{k}\psi_1\big|_{j} \times\psi_1^{-1}\big| _{k-j} \ ,
\end{equation}
where $\psi_1\big|_j$ denotes the term of order $j$ in the expansion of $\psi_1$.

Finally, multiplying $\psi_1^{-1}$ and $\psi_0$, we obtain the associator:
\begin{equation}
    \Psi=\psi_1^{-1}\psi_0=1-\frac1{\kappa^2}\frac{\pi^2}{6}[A,B]+\ldots
\end{equation}
where the ellipsis denotes higher-order terms in $1/\kappa$.
\subsection{Associators in the presence of an irregular singularity}
In the case of one rank one irregular singular point sitting at $x=\infty$, the associator is modified in the following way. We consider the equation
\begin{equation}
    \kappa\partial_x\psi=(H+\frac{A}{x}+\frac{B}{x-1})\psi.
\end{equation}
Let $\psi=1+\frac1\kappa(f_HH+f_AA+f_BB)+\frac1{\kappa^2}(f_{HA}HA+\ldots)+\ldots$. 
Again we obtain a hierarchy of differential equations by comparing the powers of $\frac1\kappa$:
\begin{equation}
    \partial_xf_H=1,\partial_xf_A=\frac1x,\partial_xf_B=\frac{1}{x-1},\partial_xf_{H^2}=f_H,\ldots
\end{equation}
We can solve these functions recursively:
\begin{equation}
    \begin{gathered}
        f_H=x+C_H,f_A=\log x+C_A,f_B=\log(1-x)+C_B,\\
        f_{H^2}=\frac12x^2+C_Hx+C_{H^2},f_{HA}=x\log x-x +C_Ax+C_{HA},\\
        f_{HB}=x\log(1-x)-\log (1-x)-x+C_Bx+C_{HB},\\
        f_{AH}=x+C_H\log x+C_{AH},f_{A^2}=\frac12(\log x)^2+C_A\log x+C_{A^2},\\f_{AB}=\mathrm{Li}_2(1-x)+\log x\log(1-x)+C_B\log x+C_{AB},\\
        f_{BH}=x+\log(1-x)+C_H\log(1-x)+C_{BH},\\
    f_{BA}=-\mathrm{Li}_2(1-x)+C_A\log(1-x)+C_{BA},\\f_{B^2}=\frac12(\log (1-x))^2+C_B\log(1-x)+C_{B^2}.
    \end{gathered}
\end{equation}
Then,
for $\psi_0\sim x^A$ around $x=0$, we have $C_H=0,C_A=0,C_B=C_{H^2}=0,C_{HA}=0,C_{HB}=0.$ $C_{AH}=C_{A^2}=C_{BH}=C_{B^2}=0$. And $C_{AB}=-\frac{\pi^2}{6},C_{BA}=\frac{\pi^2}{6}.$
\begin{equation}
\begin{gathered}
      \psi_0=1+\frac1\kappa(xH+\log x A+\log(1-x)B)+\frac1{\kappa^2}(\frac12x^2H^2+(x\log x-x)HA+\\
      (x\log (1-x)-\log(1-x)-x)HB+xAH+\frac12(\log x)^2A^2+\\
      (\mathrm{Li}_2(1-x)+\log x\log(1-x)-\frac{\pi^2}{6})AB+(x+\log(x-1))BH\\
      -\mathrm{Li}_2(1-x)BA+\frac12(\log(1-x))^2 B^2)+\ldots
\end{gathered}
\end{equation}

For $\psi_1\sim (1-x)^B$ around $x=1$, we have $C_H=-1,C_A=0,C_B=0$, $C_{HH}=\frac12,C_{HA}=0,C_{HB}=1,C_{AH}=-1,C_{AA}=0,C_{AB}=0,C_{BH}=-1,C_{BA}=0,C_{BB}=0.$
\begin{equation}
\begin{gathered}
       \psi_1=1+\frac1\kappa((x-1)H+\log x A+\log(1-x)B)+\frac{1}{\kappa^2}((\frac{x^2}{2}-x+\frac12)H^2+(x\log x-x)HA\\
       +(x\log (1-x)-\log (1-x)-x+1)HB+(x-1-\log x)AH+\frac12(\log x)^2A^2+\\
       (\mathrm{Li}_2(1-x)+\log x\log(1-x))AB+(x-1)BH\\-\mathrm{Li}_2(1-x)BA
       +\frac12(\log (1-x))^2B^2)+\ldots
\end{gathered}
\end{equation}
The associator is 
\begin{equation}
    \Psi=\psi_1^{-1}\psi_0=1+\frac1\kappa H+\frac1{\kappa^2}(\frac{1}{2}H^2+[A,H]+[B,H]-\frac{\pi^2}{6}[A,B])+\ldots
\end{equation}

\subsection{Exact solutions:the regular case}
We consider the Fuchsian system defined by the matrices
\begin{equation}
  A= \begin{pmatrix}
       -\frac32&0\\
    0&\frac12
   \end{pmatrix} , B=\begin{pmatrix}
    0&\frac{\sqrt3}{2}\\ \frac{\sqrt3}{2}&-1
\end{pmatrix}.
\end{equation}
The differential equation
\begin{equation}
    \kappa\partial_x\psi=(\frac{A}{x}+\frac{B}{x-1})\psi
\end{equation}
possesses solutions expressible in terms of the Gauss hypergeometric function $_2F_1(a,b;c;x)$ \cite{Bateman:100233}.
Let the parameters be fixed as
\begin{equation}
    a=-\frac{1}{\kappa},b=\frac{1}{\kappa},c=\frac{-2}{\kappa}.
\end{equation}
The fundamental scalar equation associated with this system is:
\begin{equation}
    x(1-x)f''(x)-(x+\frac2\kappa)f'(x)+\frac{1}{\kappa^2}f(x)=0.
\end{equation}
Near the singularity $x=0$, we define the following basis of solutions:
\begin{equation}
\begin{aligned}
     u_1&=x^{-\frac{3}{2\kappa}}(1-x)^{\frac{1}{2\kappa}} {_2}F_1(-\frac{1}{\kappa},\frac{1}{\kappa};-\frac{2}{\kappa};x),\\
     \tilde{u}_1&=\frac{x-1}{\sqrt3x}(3u_1+2\kappa xu_1'),
\end{aligned}
\end{equation}
\begin{equation}
\begin{aligned}
     u_5&=x^{1+\frac{1}{2\kappa}}(1-x)^{\frac{1}{2\kappa}} {_2}F_1(1+\frac{1}{\kappa},1+\frac{3}{\kappa};2+\frac{2}{\kappa};x),\\
     \tilde{u}_5&=\frac{x-1}{\sqrt3x}(3u_5+2\kappa xu_5').
\end{aligned}
\end{equation}
We construct a scaled solution matrix $\psi_0$ that exhibits the desired asymptotic behavior as $x\rightarrow 0$:
\begin{equation}
\psi_0=
\begin{pmatrix}
        u_1& -\frac{\sqrt3}{2(2+\kappa)}u_5\\
    \tilde{u}_1&-\frac{\sqrt3}{2(2+\kappa)}\tilde{u}_5
\end{pmatrix}
\sim\begin{pmatrix}
    1+\mathcal{O}(x)&\mathcal{O}(x)\\
    \mathcal{O}(x)&1+\mathcal{O}(x)
\end{pmatrix}\begin{pmatrix}
    x^{-\frac{3}{2\kappa}}&0\\
    0&x^{\frac{1}{2\kappa}}
\end{pmatrix}.
\end{equation}
Similarly, near $x=1$, the relevant basis solutions are:
\begin{equation}
    \begin{gathered}
        u_2=x^{-\frac{3}{2\kappa}}(1-x)^{\frac{1}{2\kappa}} {_2}F_1(1+\frac1\kappa,1+\frac3\kappa;1+\frac{2}{\kappa};1-x),\\
    \tilde{u}_2=\frac{x-1}{\sqrt3x}(3u_2+2\kappa x u'_2) ,
    \end{gathered}
\end{equation}
\begin{equation}
\begin{gathered}
       u_6= x^{\frac{-3}{2\kappa}}(1-x)^{-\frac{3}{2\kappa}} {_2}F_1(-\frac1\kappa,-\frac3\kappa;1-\frac2\kappa;1-x),\\
\tilde{u}_6=\frac{x-1}{\sqrt3x}(3u_6+2\kappa x u'_6) .      
\end{gathered}
\end{equation}
We define a solution matrix $\psi_1$ normalized to satisfy the asymptotic behavior $\sim(1-x)^B$:
\begin{equation}
   \psi_1= \begin{pmatrix}
        \frac14u_6+\frac{3}{4}u_2&-\frac{\sqrt3}{4}u_6+\frac{\sqrt3}{4}u_2\\
         \frac14\tilde u_6+\frac{3}{4}\tilde u_2&-\frac{\sqrt3}{4}\tilde u_6+\frac{\sqrt3}{4}\tilde u_2 
    \end{pmatrix}.
\end{equation}
We have the relation:
 \begin{equation}
    \begin{pmatrix}
        u_2&u_6
    \end{pmatrix}\begin{pmatrix}
        \frac{\Gamma(c)\Gamma(c-a-b)}{\Gamma(c-a)\Gamma(c-b)}&\frac{\Gamma(2-c)\Gamma(c-a-b)}{\Gamma(1-a)\Gamma(1-b)}\\
        \frac{\Gamma(c)\Gamma(a+b-c)}{\Gamma(a)\Gamma(b)}&\frac{\Gamma(2-c)\Gamma(a+b-c)}{\Gamma(a+1-c)\Gamma(b+1-c)}
    \end{pmatrix}=
    \begin{pmatrix}
        u_1&u_5
    \end{pmatrix}
\end{equation} 
So $\Psi$ is given by:
\begin{equation}
\begin{gathered}
       \psi_0=\psi_1 \Psi,\\
\Psi=\begin{pmatrix}
    1&1\\
    -\sqrt3&\frac{1}{\sqrt3}
\end{pmatrix}       \begin{pmatrix}
     \frac{\Gamma(c)\Gamma(a+b-c)}{\Gamma(a)\Gamma(b)}&\frac{-\sqrt3}{2(2+\kappa)}\frac{\Gamma(2-c)\Gamma(a+b-c)}{\Gamma(a+1-c)\Gamma(b+1-c)}   \\
         \frac{\Gamma(c)\Gamma(c-a-b)}{\Gamma(c-a)\Gamma(c-b)}&\frac{-\sqrt3}{2(2+\kappa)}\frac{\Gamma(2-c)\Gamma(c-a-b)}{\Gamma(1-a)\Gamma(1-b)}
    \end{pmatrix}.
\end{gathered}
\end{equation}
Expanding $\Psi$ as a power series in $\kappa^{-1}$, we obtain:
\begin{equation}
\begin{gathered}
     \Psi=\begin{pmatrix}
        1&0\\
        0&1
    \end{pmatrix}+\frac{1}{\kappa^2}\begin{pmatrix}
        0&\frac{\pi^2}{6}\sqrt3\\
        -\frac{\pi^2}{6}\sqrt3&0
    \end{pmatrix}+\\
    \frac{1}{\kappa^3}\begin{pmatrix}
         \frac32\uppsi_2(1)&\frac{\sqrt3}{2}\uppsi_2(1)\\
        \frac{\sqrt3}{2}\uppsi_2(1)&\frac{8\uppsi_2(2)-17\uppsi_2(1)-16}{6}
    \end{pmatrix}+\mathcal{O}(\frac{1}{\kappa^4}).
\end{gathered}
\end{equation}
The first two terms coincide with the expansion $1+\frac{\pi^2}{6}[B,A]$. To relate higher-order terms to the Drinfeld associator, we use the polygamma function $\uppsi_n(z):=\frac{d^{m+1}}{dz^{m+1}}\log \Gamma(z)$ and its values at integer arguments~\cite{abramowitz+stegun};
\begin{equation}
\begin{aligned}
        \uppsi_m(1)&=(-1)^{m+1}m!\zeta(m+1),\\
        \uppsi_m(n+1)&=(-1)^{m+1}m![\zeta(m+1)-1-\frac{1}{2^{m+1}}-\ldots-\frac{1}{n^{m+1}}].
\end{aligned}
\end{equation}
\subsection{Integral representations}
\subsubsection{$\Lambda=0$ case}
Let
\begin{equation}
    g=[w(w-1)(w-z)]^{1/\kappa}
\end{equation}
and
\begin{equation}
    f=-\frac1\kappa\begin{pmatrix}
        \frac{g}{w-1}\\
        \frac{g}{w-z}\\
        \frac{g}{w}
    \end{pmatrix}.
\end{equation}
The following equation
\begin{equation}
    \kappa\partial_z\psi=\left[\frac{1}{z}\begin{pmatrix}
        0&0&0\\
        0&1&-1\\
        0&-1&1
    \end{pmatrix}+\frac{1}{z-1}\begin{pmatrix}
        1&-1&0\\
        -1&1&0\\
        0&0&0
    \end{pmatrix}\right]\psi
\end{equation}
is solved by $\psi_1=\int_0^z f dw$ and $\psi_2=\int_z^1 f dw$ in certain value range of $\kappa$ and $z$. The third linearly independent solution is $\psi_3=\begin{pmatrix}
    1\\1\\1
\end{pmatrix}$, a constant vector.
Explicitly, we have:
\begin{equation}
    \psi_1=\begin{pmatrix}
        -z^{2/\kappa+1}B(1+\frac1\kappa,1+\frac1\kappa ){}_2F_1(1+\frac1\kappa,1-\frac1\kappa,2+\frac2\kappa,z)\\
        -z^{2/\kappa}B(1+\frac1\kappa,\frac1\kappa ){}_2F_1(1+\frac1\kappa,-\frac1\kappa,1+\frac2\kappa,z)\\z^{2/\kappa}B(1+\frac1\kappa,\frac1\kappa ){}_2F_1(-\frac1\kappa,\frac1\kappa,1+\frac2\kappa,z)
    \end{pmatrix},
\end{equation}

\begin{equation}
    \psi_2=\begin{pmatrix}
-2B(1+\frac1\kappa,\frac2\kappa){}_2F_1(-\frac3\kappa,-\frac1\kappa,-\frac2\kappa,z)-z^{\frac2\kappa+1}B(1+\frac1\kappa,-1-\frac2\kappa){}_2F_1(1+\frac1\kappa,1-\frac1\kappa,2+\frac2\kappa,z)\\
        B(1+\frac1\kappa,\frac2\kappa){}_2F_1(-\frac3\kappa,1-\frac1\kappa,1-\frac2\kappa,z)+z^{\frac2\kappa}B(-\frac2\kappa,\frac1\kappa){}_2F_1(1+\frac1\kappa,-\frac1\kappa,1+\frac2\kappa,z)\\
        B(1+\frac1\kappa,\frac2\kappa){}_2F_1(-\frac3\kappa,-\frac1\kappa,1-\frac2\kappa,z)-z^{\frac2\kappa}B(-\frac2\kappa,\frac1\kappa){}_2F_1(-\frac1\kappa,\frac1\kappa,1+\frac2\kappa,z)
    \end{pmatrix},
\end{equation}
where $B(\cdot,\cdot)$ is the beta function.
By series expansion the behavior of $(\psi_3,\psi_2+\frac12\sec(\frac\pi\kappa)\psi_1,\psi_1)$ around $z=0$ is 
\begin{equation}
    \begin{pmatrix}
        1&-2B(1+\frac1\kappa,\frac2\kappa)+\mathcal{O}(z)&\mathcal{O}(z)\\
        1&B(1+\frac1\kappa,\frac2\kappa)&-B(\frac1\kappa,1+\frac1\kappa)z^{\frac2\kappa}\\
         1&B(1+\frac1\kappa,\frac2\kappa)&B(\frac1\kappa,1+\frac1\kappa)z^{\frac2\kappa}
    \end{pmatrix}.
\end{equation}

Using the relation between $u_1,u_5$ and $u_2,u_6$, we can write down the behavior of $(\psi_3,\psi_2,\psi_1+\frac12\sec(\frac\pi\kappa)\psi_2)$ around $z=1$:
\begin{equation}
    \begin{pmatrix}
         1&2\cos(\frac\pi\kappa)B(1+\frac1\kappa,-\frac2\kappa)(1-z)^{\frac2\kappa}&-B(1+\frac1\kappa,\frac2\kappa)\\
        1& -2\cos(\frac\pi\kappa)B(1+\frac1\kappa,-\frac2\kappa)(1-z)^{\frac2\kappa}&-B(1+\frac1\kappa,\frac2\kappa)\\
        1& \mathcal{O}(1-z)&B(\frac1\kappa,1+\frac2\kappa)
    \end{pmatrix}.
\end{equation}

Now we have bases of fundamental solution matrix. We should arrange the leading term according to the behavior around $z=0$:
\begin{equation}
    z^A=\begin{pmatrix}
        1&0&0\\
        0&\frac{1+x^\frac2\kappa}{2}&\frac{1-x^\frac2\kappa}{2}\\
        0&\frac{1-x^\frac2\kappa}{2}&\frac{1+x^\frac2\kappa}{2}
        \end{pmatrix}.
\end{equation}
Let us denote the solution that has this behavior as $G_0$. We can express $G_0$ using $(\psi_1,\psi_2,\psi_3)$ and a matrix $F_0$:
\begin{equation}
\begin{aligned}
    G_0&=(\psi_1,\psi_2,\psi_3)F_0,\\
    F_0&=\begin{pmatrix}
        \frac{\alpha y}{xY-yX}&\frac12(-\frac{\alpha Y}{xY-yX}-\frac1t)&\frac12(-\frac{\alpha Y}{xY-yX}+\frac1t)\\
        \frac{- y}{xY-yX}&\frac12\frac{ Y}{xY-yX}&\frac12\frac{ Y}{xY-yX}\\
        \frac{x}{xY-yX}&-\frac12\frac{X}{xY-yX}&\frac12\frac{X}{xY-yX}
    \end{pmatrix}.
\end{aligned}
\end{equation}
Here $\alpha=-\frac12\sec{\frac\pi\kappa}$, $Y=y=1$.

The $(1-z)^B$ has the following form:
   \begin{equation}
       \begin{pmatrix}
           \frac{1+(1-z)^\frac2\kappa}{2}&\frac{1-(1-z)^\frac2\kappa}{2}&0\\
           \frac{1-(1-z)^\frac2\kappa}{2}&\frac{1+(1-z)^\frac2\kappa}{2}&0\\
           0&0&1
       \end{pmatrix}.
   \end{equation}
We denote the solution of this behavior as $G_1$. Similarly we have 
\begin{equation}
    G_1=(\psi_1,\psi_2,\psi_3) F_1
\end{equation}
where $F_1$ similarly can be written as matrix of $s,X',x$ and $\alpha$.
The associator we want is 
\begin{equation}
    \begin{aligned}
        \Psi_0&=G_1^{-1}G_0\\
        &=F_1^{-1}(\psi_1,\psi_2,\psi_3)^{-1}(\psi_1,\psi_2,\psi_3)F_0\\
        &=F_1^{-1}F_0
    \end{aligned}.
\end{equation}
Again, one can verify(though we won't do it explicitly here) that the associator expands to 
\begin{equation}
\Psi_{\Lambda=0}(1/\kappa)=1+\frac{\pi^2}{6}[B,A]+\mathcal{O}(\frac1{\kappa^3}).
\end{equation}
\subsubsection{Nonzero $\Lambda$ }
Consider 
\begin{equation}
    g=[w(w-1)(w-z)]^{1/\kappa}\exp(-\frac{\Lambda}{\kappa}w)
\end{equation}
and
\begin{equation}
    \psi=-\frac1\kappa\begin{pmatrix}
        \int_{\Gamma}\frac{g}{w-1}\\
        \int_\Gamma\frac{g}{w-z}\\
        \int_\Gamma\frac{g}{w}
    \end{pmatrix}.
\end{equation}
$\psi$ solves the following equation
\begin{equation}
    \kappa\partial_z\psi=(\frac{1}{z}\begin{pmatrix}
        0&0&0\\
        0&1&-1\\
        0&-1&1
    \end{pmatrix}+\frac{1}{z-1}\begin{pmatrix}
        1&-1&0\\
        -1&1&0\\
        0&0&0
    \end{pmatrix}+
    \Lambda\begin{pmatrix}
        0&0&0\\
        0&-1&0\\
        0&0&0
    \end{pmatrix})\psi
\end{equation}
when we choose the integration contour $\Gamma$ such that $\frac{g}{w-z}$ vanish at the two end points of $\Gamma$.

For $z<<0$, the $\Lambda$ effect on $\psi_1$ is small. Then we have the approximation
\begin{equation}
    \begin{aligned}
        \psi_1&=\int_{0}^z\begin{pmatrix}
            w^\frac1\kappa(w-z)^\frac{1}{\kappa}(w-1)^{\frac{1}{\kappa}-1}\exp(-\frac{\Lambda}{\kappa}w)\\
            w^\frac{1}{\kappa}(w-z)^{\frac{1}{\kappa}-1}(w-1)^{\frac{1}{\kappa}}\exp(-\frac{\Lambda}{\kappa}w)\\
            w^{\frac{1}{\kappa}-1}(w-z)^{\frac{1}{\kappa}}(w-1)^{\frac{1}{\kappa}}\exp(-\frac{\Lambda}{\kappa}w)
        \end{pmatrix}dw\\
        &\sim\begin{pmatrix}
           \mathcal{O}(z)\\
            -B(\frac1\kappa,1+\frac1\kappa)z^{\frac{2}{\kappa}}=-tz^{\frac2\kappa}\\B(\frac1\kappa,1+\frac1\kappa)z^{\frac{2}{\kappa}}=tz^{\frac{2}{\kappa}}
        \end{pmatrix}.
    \end{aligned}
\end{equation}
Hereafter we assume that the non-holomorphic piece ($z^{2/\kappa},(1-z)^{2/\kappa}$) in expansion is not affected by nonzero $\Lambda$.
\begin{equation}
   \begin{aligned}
        \psi_2&=\int_{z}^1\begin{pmatrix}
            w^\frac1\kappa(w-z)^\frac{1}{\kappa}(w-1)^{\frac{1}{\kappa}-1}\exp(-\frac{\Lambda}{\kappa}w)\\
            w^\frac{1}{\kappa}(w-z)^{\frac{1}{\kappa}-1}(w-1)^{\frac{1}{\kappa}}\exp(-\frac{\Lambda}{\kappa}w)\\
            w^{\frac{1}{\kappa}-1}(w-z)^{\frac{1}{\kappa}}(w-1)^{\frac{1}{\kappa}}\exp(-\frac{\Lambda}{\kappa}w)
        \end{pmatrix}dw\\
        &\sim \int_0^1\begin{pmatrix}
            w^\frac2\kappa(w-1)^{\frac{1}{\kappa}-1}\exp(-\frac{\Lambda}{\kappa}w)\\
            w^{\frac{2}{\kappa}-1}(w-1)^{\frac{1}{\kappa}}\exp(-\frac{\Lambda}{\kappa}w)\\
            w^{\frac{2}{\kappa}-1}(w-1)^{\frac{1}{\kappa}}\exp(-\frac{\Lambda}{\kappa}w)
        \end{pmatrix}dw+z^{2/\kappa} \text{ terms} \\
        &=\begin{pmatrix}
           -2B(\frac1\kappa+1,\frac2\kappa){}_1F_1(\frac{2}{\kappa}+1,\frac{3}{\kappa}+1,-\frac{\Lambda}{\kappa})=X\\
            B(\frac1\kappa+1,\frac2\kappa){}_1F_1(\frac{2}{\kappa},\frac{3}{\kappa}+1,-\frac{\Lambda}{\kappa})=x\\
            B(\frac1\kappa+1,\frac2\kappa) {}_1F_1(\frac{2}{\kappa},\frac{3}{\kappa}+1,-\frac{\Lambda}{\kappa})=x
        \end{pmatrix}\\
        &+\begin{pmatrix}
            \mathcal{O}(z)\\
           B(-\frac2\kappa,\frac1\kappa)
           \\
           -B(-\frac2\kappa,\frac1\kappa)
        \end{pmatrix}z^{2/\kappa}
    \end{aligned},
\end{equation}

\begin{equation}
    \begin{aligned}
\psi_3&=\frac{(\frac\Lambda\kappa)^{\frac3\kappa}}{\Gamma(\frac3\kappa)}\int_{1}^{\infty}\begin{pmatrix}
            w^\frac1\kappa(w-z)^\frac{1}{\kappa}(w-1)^{\frac{1}{\kappa}-1}\exp(-\frac{\Lambda}{\kappa}w)\\
            w^\frac{1}{\kappa}(w-z)^{\frac{1}{\kappa}-1}(w-1)^{\frac{1}{\kappa}}\exp(-\frac{\Lambda}{\kappa}w)\\
            w^{\frac{1}{\kappa}-1}(w-z)^{\frac{1}{\kappa}}(w-1)^{\frac{1}{\kappa}}\exp(-\frac{\Lambda}{\kappa}w)
        \end{pmatrix}dw\\
        &\sim \frac{(\frac\Lambda\kappa)^{\frac3\kappa}}{\Gamma(\frac3\kappa)} \int_1^\infty\begin{pmatrix}
            w^\frac2\kappa(w-1)^{\frac{1}{\kappa}-1}\exp(-\frac{\Lambda}{\kappa}w)\\
            w^{\frac{2}{\kappa}-1}(w-1)^{\frac{1}{\kappa}}\exp(-\frac{\Lambda}{\kappa}w)\\
            w^{\frac{2}{\kappa}-1}(w-1)^{\frac{1}{\kappa}}\exp(-\frac{\Lambda}{\kappa}w)
        \end{pmatrix}dw\\
        &=\begin{pmatrix}
            (\frac\Lambda\kappa)^{\frac3\kappa}\frac{\Gamma(-\frac
            3\kappa)\Gamma(\frac1\kappa)}{\Gamma(\frac3\kappa)\Gamma(-\frac2\kappa)}{}_1F_1(\frac2\kappa+1,\frac3\kappa+1,-\frac\Lambda\kappa)+{}_1F_1(1-\frac1\kappa,1-\frac3\kappa,-\frac\Lambda\kappa)\\
           (\frac\Lambda\kappa)^{\frac3\kappa} \frac{\Gamma(-\frac
            3\kappa)\Gamma(\frac1\kappa+1)}{\Gamma(\frac3\kappa)\Gamma(1-\frac2\kappa)}{}_1F_1(\frac2\kappa,\frac3\kappa+1,-\frac\Lambda\kappa)+{}_1F_1(-\frac1\kappa,1-\frac3\kappa,-\frac\Lambda\kappa)\\
            (\frac\Lambda\kappa)^{\frac3\kappa}\frac{\Gamma(-\frac
            3\kappa)\Gamma(\frac1\kappa+1)}{\Gamma(\frac3\kappa)\Gamma(1-\frac2\kappa)}{}_1F_1(\frac2\kappa,\frac3\kappa+1,-\frac\Lambda\kappa)+{}_1F_1(-\frac1\kappa,1-\frac3\kappa,-\frac\Lambda\kappa)
        \end{pmatrix}\\
        &=\frac{(\frac\Lambda\kappa)^{\frac3\kappa}}{\Gamma(\frac3\kappa)}\begin{pmatrix}
             e^{-\frac{\Lambda}{\kappa}}\Gamma(\frac1\kappa)U(\frac1\kappa,\frac3\kappa+1,\frac{\Lambda}{\kappa})\\
            e^{-\frac{\Lambda}{\kappa}}\Gamma(\frac1\kappa+1)U(\frac1\kappa+1,\frac3\kappa+1,\frac{\Lambda}{\kappa})\\
             e^{-\frac{\Lambda}{\kappa}}\Gamma(\frac1\kappa+1)U(\frac1\kappa+1,\frac3\kappa+1,\frac{\Lambda}{\kappa})
        \end{pmatrix}\sim\begin{pmatrix}
            1\\1\\1
        \end{pmatrix}
    \end{aligned}.
\end{equation}
The following solution matrix has a relatively simple asymptotic behavior:
\begin{equation}
\begin{gathered}
\begin{pmatrix}
        \psi_3-(\frac\Lambda\kappa)^{\frac3\kappa}\frac{\Gamma(1-\frac3\kappa)}{\Gamma(1-\frac2\kappa)\Gamma(\frac2\kappa)}(\psi_2+\frac12\sec(\frac\pi\kappa)\psi_1),&\psi_2+\frac12\sec(\frac\pi\kappa)\psi_1,&\psi_1
    \end{pmatrix}\\
   \sim \begin{pmatrix}
        {}_1F_1(1-\frac1\kappa,1-\frac3\kappa,-\frac\Lambda\kappa)=Y&X&0\\
 {}_1F_1(-\frac1\kappa,1-\frac3\kappa,-\frac\Lambda\kappa)=y&x&-tz^{\frac2\kappa}\\
  {}_1F_1(-\frac1\kappa,1-\frac3\kappa,-\frac\Lambda\kappa)=y&x&tz^{\frac2\kappa}
    \end{pmatrix}
\end{gathered}
\end{equation}
One can easily obtain the desired solution by linear combinations and scaling of numbers $X,Y,x,y,t$.
Now we write the solution around $z=1$:
\begin{equation}
   \begin{aligned}
       \psi_2=\int_z^1\ldots
       \\
       =\exp(-\frac\Lambda\kappa)&\begin{pmatrix}
        2\cos(\frac\pi\kappa)B(-\frac2\kappa,\frac1\kappa+1)(1-z)^\frac2\kappa=-s(1-z)^\frac2\kappa\\
        -2\cos(\frac\pi\kappa)B(-\frac2\kappa,\frac1\kappa+1)(1-z)^\frac2\kappa=s(1-z)^\frac2\kappa\\
        O((1-z)^{\frac2\kappa})
    \end{pmatrix} , \\
    \psi_1+\frac12\sec(\frac\pi\kappa)\psi_2&=\begin{pmatrix}
        -B(1+\frac1\kappa,\frac2\kappa){}_1F_1(\frac1\kappa+1,\frac3\kappa+1,-\frac\Lambda\kappa)=X'\\
            -B(1+\frac1\kappa,\frac2\kappa){}_1F_1(\frac1\kappa+1,\frac3\kappa+1,-\frac\Lambda\kappa)=X'\\
            2B(1+\frac1\kappa,\frac2\kappa){}_1F_1(\frac1\kappa,\frac3\kappa+1,-\frac\Lambda\kappa)=x'
        \end{pmatrix},\\
   \end{aligned}
   \end{equation}
\begin{equation}
    \begin{aligned}
        \psi_3&=\frac{(\frac\Lambda\kappa)^{\frac3\kappa}}{\Gamma(\frac3\kappa)}\int_1^\infty\cdots\\
        &\sim\begin{pmatrix}
            (\frac\Lambda\kappa)^{\frac3\kappa}\frac{\Gamma(-\frac3\kappa)\Gamma(\frac2\kappa)}{\Gamma(-\frac1\kappa)\Gamma(\frac3\kappa)}{}_1F_1(1+\frac1\kappa,1+\frac3\kappa,-\frac\Lambda\kappa)+{}_1F_1(1-\frac2\kappa,1-\frac3\kappa,-\frac\Lambda\kappa)
            \\
            (\frac\Lambda\kappa)^{\frac3\kappa}\frac{\Gamma(-\frac3\kappa)\Gamma(\frac2\kappa)}{\Gamma(-\frac1\kappa)\Gamma(\frac3\kappa)}{}_1F_1(1+\frac1\kappa,1+\frac3\kappa,-\frac\Lambda\kappa)+{}_1F_1(1-\frac2\kappa,1-\frac3\kappa,-\frac\Lambda\kappa)\\
            (\frac\Lambda\kappa)^{\frac3\kappa}\frac{\Gamma(-\frac3\kappa)\Gamma(1+\frac2\kappa)}{\Gamma(1-\frac1\kappa)\Gamma(\frac3\kappa)}{}_1F_1(\frac1\kappa,1+\frac3\kappa,-\frac\Lambda\kappa)+{}_1F_1(-\frac2\kappa,1-\frac3\kappa,-\frac\Lambda\kappa)
        \end{pmatrix}\\
        &= \frac{(\frac\Lambda\kappa)^{\frac3\kappa}}{\Gamma(\frac3\kappa)}\exp(-\frac\Lambda\kappa)\begin{pmatrix}
           \Gamma(\frac2\kappa)U(\frac2\kappa,\frac3\kappa+1,\frac\Lambda\kappa)\\\Gamma(\frac2\kappa)U(\frac2\kappa,\frac3\kappa+1,\frac\Lambda\kappa)\\
\Gamma(\frac2\kappa+1)U(\frac2\kappa+1,\frac3\kappa+1,\frac\Lambda\kappa)
        \end{pmatrix}.
    \end{aligned}
\end{equation}
The following solution matrix has a relatively simple behavior:
\begin{equation}
    \begin{gathered}
            \begin{pmatrix}
        \psi_3-(\frac\Lambda\kappa)^{\frac3\kappa}\frac{\Gamma(1-\frac3\kappa)}{\Gamma(1-\frac1\kappa)\Gamma(\frac1\kappa)}(\psi_1+\frac12\sec(\frac\pi\kappa)\psi_2),&\psi_1+\frac12\sec(\frac\pi\kappa)\psi_2,&\psi_2
    \end{pmatrix}
    \\
    \sim\begin{pmatrix}
        {}_1F_1(1-\frac2\kappa,1-\frac3\kappa,-\frac\Lambda\kappa)=Y'&X'&-s(1-z)^{\frac2\kappa}\\
 {}_1F_1(1-\frac2\kappa,1-\frac3\kappa,-\frac\Lambda\kappa)=Y'&X'&s(1-z)^{\frac2\kappa}\\
 {}_1F_1(-\frac2\kappa,1-\frac3\kappa,-\frac\Lambda\kappa)=y'&x'&0
    \end{pmatrix}
    \end{gathered}.
\end{equation}
Again, the standard solution can be obtained by manipulating numbers $Y',X',y',x'$ and $s$.

The explicit form of the associator in terms of these quantities($X,X', etc$) is intricate, and we omit it here. It suffices to note that $\tilde{\Psi}_\Lambda$ admits the following expansion in $\Lambda$:
\begin{equation}
  \begin{gathered}
\tilde{\Psi}(\Lambda,1/\kappa)\sim \hat{\Psi}+\Psi,\\
\Psi=\Psi_{0}(1/\kappa)+\Lambda\Psi_1(1/\kappa)+\Lambda^2\Psi_2(1/\kappa)+\ldots\\
\hat{\Psi}=(\frac{\Lambda}{\kappa})^{\frac3\kappa}(\hat{\Psi}_0(1/\kappa)+\Lambda\hat{\Psi}_1(1/\kappa)+\Lambda^2\hat{\Psi}_2(1/\kappa)+\ldots)
\end{gathered}
\end{equation}
where one can verify that $\Psi_0(1/\kappa)$ matches the associator in the $\Lambda=0$ case. Also, by expansion in $\frac1\kappa$ one can verify
\begin{equation}
    \Psi_1(1/\kappa)=H+[A,H]+[B,H]+\ldots
\end{equation}
and so on. 

The local monodromy and braiding matrices are 
\begin{equation}
\begin{aligned}
     M_A\doteq\begin{pmatrix}
        1&0&0\\
        0&\frac{1+q^2}2&\frac{1-q^2}{2}\\
     0&\frac{1-q^2}2&\frac{1+q^2}{2}
    \end{pmatrix},\\
    B_A\doteq\begin{pmatrix}
        1&0&0\\
        0&\frac{1-q^2}2&\frac{1+q^2}{2}\\
     0&\frac{1+q^2}2&\frac{1-q^2}{2}
    \end{pmatrix},\\
    M_B\doteq\begin{pmatrix}
        \frac{1+q^2}2&\frac{1-q^2}{2}&0\\
     \frac{1-q^2}2&\frac{1+q^2}{2}&0\\
     0&0&1
    \end{pmatrix},\\
    B_B\doteq\begin{pmatrix}
        \frac{1-q^2}2&\frac{1+q^2}{2}&0\\
     \frac{1+q^2}2&\frac{1-q^2}{2}&0\\
     0&0&1
    \end{pmatrix}.
\end{aligned}
\end{equation}
Some examples of product of matrices associated to braids:
\begin{equation}
\begin{aligned}
\Psi_\Lambda M_A\Psi_\Lambda^{-1}\\
     \Psi_\Lambda, M_A\Psi_\Lambda^{-1}M_B\\
     \Psi_\Lambda, B_A\Psi_\Lambda^{-1}B_B.
\end{aligned}
\end{equation}
The trace of these matrix products are independent of $\Lambda$. In particular, $F_0 M_A(B_A)F_0^{-1}$ and $F_1M_B(B_B)F_1^{-1}$ are independent of $\Lambda$. This confirms the independence of monodromy on the value of $\Lambda$ in the single irregular singularity case, as we discussed in section 3.
\bibliographystyle{JHEP}     
{\small{\bibliography{references}}}
\end{document}